\documentclass[11pt,a4paper,twoside,fleqn]{article}
\usepackage{color,graphicx,natbib,lscape,here,geometry,setspace,multirow,enumitem}
\usepackage[dvipsnames]{xcolor}
\usepackage{graphicx}
\usepackage{authblk}
\usepackage{silence}
\usepackage{rotating}
\usepackage{multirow}

\usepackage{booktabs}
\usepackage{diagbox}
\usepackage{color}
\usepackage{setspace}
\usepackage{algorithm2e}
\usepackage{enumitem}
\usepackage{tikz}

\definecolor{nblue}{HTML}{000660}
\usepackage[colorlinks=true,urlcolor=nblue,linkcolor=nblue,citecolor=nblue]{hyperref}
\usepackage{bigstrut}

\usepackage{tabularx}
\usepackage{threeparttable}
\usepackage{dcolumn}
\newcolumntype{d}[1]{D{.}{.}{#1}}

\usepackage{etoolbox} 
\makeatletter
\patchcmd{\BR@backref}{\newblock}{\newblock[}{}{}
\patchcmd{\BR@backref}{\par}{]\par}{}{}
\makeatother

\usepackage{pdflscape}
\usepackage{afterpage}

\usepackage{longtable}
\usepackage{array}
\newcolumntype{C}[1]{>{\centering\arraybackslash}p{#1}}

\usepackage{comment}
\usepackage{mathtools}

\usepackage[hang,flushmargin]{footmisc} 

\usepackage[hang]{footmisc}
\usepackage[british]{babel}

\pdfoutput=1
 \usepackage{float}
 \restylefloat{table}

\usepackage[title,titletoc]{appendix}
\makeatletter

\renewenvironment{appendices}{%
    \begin{oldappendices}%
    \renewcommand{\thefigure}{\ifnum \c@section>\z@ \thesection.\fi\@arabic\c@figure}%
    \@addtoreset{figure}{section}%
    \renewcommand{\thetable}{\ifnum \c@section>\z@ \thesection.\fi\@arabic\c@table}%
    \@addtoreset{table}{section}}{%
    \end{oldappendices}%
}\makeatother

\usepackage{titlesec} 
\titleformat{\section}[block]{\large}{\thesection. }{0em}{\MakeUppercase} 
\titleformat{\subsection}[block]{\large}{\thesubsection. }{0em}{\itshape} 
\titleformat{\subsubsection}[block]{\large}{}{0em}{\itshape} 

\let\natbibcitet\citet
\renewcommand\citet{\bibpunct{(}{)}{,}{a}{,}{,}\natbibcitet}
\let\natbibcitep\citep
\renewcommand\citep{\bibpunct{(}{)}{;}{a}{,}{;}\natbibcitep}
\newcommand{\bi}{\begin{itemize}}
\newcommand{\ei}{\end{itemize}}
\newcommand{\be}{\begin{equation}}
\newcommand{\ee}{\end{equation}}
\defcitealias{ieo14}{IEO, 2014}

\long\def\symbolfootnote[#1]#2{\begingroup%
\def\thefootnote{\fnsymbol{footnote}}\footnote[#1]{#2}\endgroup}

\makeatletter
\def\ubar#1{\underline{\sbox\tw@{$#1$}\dp\tw@\z@\box\tw@}}
\def\obar#1{\overline{\sbox\tw@{$#1$}\dp\tw@\z@\box\tw@}}
\makeatother

\usepackage{bm}
\usepackage{caption}
\usepackage{subcaption}

\makeatletter\let\p@subfigure\thefigure\makeatother

\floatstyle{plaintop}
\restylefloat{table}

\usepackage{fancyhdr} 
\usepackage{cleveref}
\crefname{chapter}{Chapter}{Chapters}
\crefname{section}{Section}{Sections}
\crefname{subsection}{Section}{Sections}
\crefname{subsubsection}{Section}{Sections}
\crefname{figure}{Figure}{Figures}
\crefname{table}{Table}{Tables}
\crefname{equation}{Equation}{Equations}
\crefname{appendix}{Appendix}{Appendices}
\crefname{appendices}{Appendix}{Appendices}

\crefname{appsec}{Appendix}{Appendices}

\usepackage{catoptions}
\makeatletter

\def\Autoref#1{%
  \begingroup
  \edef\reserved@a{\cpttrimspaces{#1}}%
  \ifcsndefTF{r@#1}{%
    \xaftercsname{\expandafter\testreftype\@fourthoffive}
      {r@\reserved@a}.\\{#1}%
  }{%
    \ref{#1}%
  }%
  \endgroup
}
\def\testreftype#1.#2\\#3{%
  \ifcsndefTF{#1autorefname}{%
    \def\reserved@a##1##2\@nil{%
      \uppercase{\def\ref@name{##1}}%
      \csn@edef{#1autorefname}{\ref@name##2}%
      \autoref{#3}%
    }%
    \reserved@a#1\@nil
  }{%
    \autoref{#3}%
  }%
}
\makeatother

\title{\LARGE{\textbf{Bayesian state-space modeling for analyzing heterogeneous network effects of US monetary policy}}}
\author{\large{
\uppercase{Niko Hauzenberger} and 
\uppercase{Michael Pfarrhofer}}\thanks{
\noindent Corresponding author: Michael Pfarrhofer. Salzburg Centre of European Union Studies, University of Salzburg. \textit{Address}: M\"{o}nchsberg 2a, 5020 Salzburg, Austria. \textit{Email}: \href{mailto:michael.pfarrhofer@sbg.ac.at}{michael.pfarrhofer@sbg.ac.at}. \textit{Phone}: +43 662 8044 3772. We thank Marco Del Negro, Manfred M. Fischer, Sylvia Fr\"uhwirth-Schnatter, Florian Huber and Thomas Reutterer for valuable comments and suggestions; and we are particularly grateful to Michael Weber for providing us with the dataset. We gratefully acknowledge financial support from the Austrian Science Fund (FWF, grant no. ZK 35) and the Oesterreichische Nationalbank (OeNB, Anniversary Fund, project no. 18127).}
\\\vspace*{-0.5em}
\textit{University of Salzburg}}
\date{}


\def\equationautorefname~#1\null{%
  Eq.~(#1)\null
}
\def\equationautorefname~#1\null{
Eq.~(#1)\null
}

\usepackage[utf8, latin1]{inputenc}                 

\begin{document}
\maketitle\thispagestyle{empty}\normalsize\vspace*{-2em}\small

\begin{center}
\begin{minipage}{0.8\textwidth}
\noindent\small Understanding disaggregate channels in the transmission of monetary policy is of crucial importance for effectively implementing policy measures. We extend the empirical econometric literature on the role of production networks in the propagation of shocks along two dimensions. First, we allow for industry-specific responses that vary over time, reflecting non-linearities and cross-sectional heterogeneities in direct transmission channels. Second, we allow for time-varying network structures and dependence. This feature captures both variation in the structure of the production network, but also differences in cross-industry demand elasticities. We find that impacts vary substantially over time and the cross-section. Higher-order effects appear to be particularly important in periods of economic and financial uncertainty, often coinciding with tight credit market conditions and financial stress. Differentials in industry-specific responses can be explained by how close the respective industries are to end-consumers.
\end{minipage}
\end{center}

\onehalfspacing\normalsize\renewcommand{\thepage}{\arabic{page}}
\newpage\normalsize

\section{Introduction}\label{sec:introduction}
A growing number of papers explores how shocks on the micro and macro level propagate through economic networks and how such shocks relate to aggregate fluctuations \citep[see, for instance,][]{gabaix2011granular,acemoglu2012network,carvalho2013great,elliott2014financial,acemoglu2015systemic,baqaee2019macroeconomic}. We contribute to this literature by analyzing the transmission of monetary policy shocks through the granular US production network. Our interest centers on assessing time-variation in the strength of network dependencies, and effects of monetary policy shocks on industry-level returns that are allowed to vary over time and the cross-section. 

Our approach relates to \citet{ozdagli2017monetary}, who generalize the setup proposed in \citet{bernanke2005explains} and \citet{gurkaynak2005actions} for analyzing the impact of changes in monetary policy on equity prices.\footnote{These articles are among a larger body of diverse literature focusing on measuring monetary non-neturality using high-frequency market surprises around central bank policy announcements \citep[see][]{cook1989effect,thorbecke1997stock,kuttner2001monetary,cochrane2002fed,rigobon2004impact,gurkaynak2005actions,gertler2015monetary,lucca2015pre,neuhierl2018monetary,nakamura2018high,altavilla2019measuring,jarocinski2018deconstructing,paul2019time}.} While \citet{bernanke2005explains} and \citet{gurkaynak2005actions} identify a significant and substantial impact of monetary surprises on aggregate stock market indices, \citet{ozdagli2017monetary} decompose these estimates into direct effects and spillovers through the production network. They use a conventional network panel model with homogenous parameters, and provide evidence for significant higher-order effects of monetary policy on stock market returns between $55$ and $85$ percent using disaggregate data on the industry-level. 

These higher-order dynamics originate from cross-industry demand elasticities to the same shock, amplifying direct effects of monetary policy interventions in the interconnected US production network. We provide extensions from an econometric and empirical perspective by drawing from the vast literature on Bayesian state-space modeling \citep[see][]{kim1999state}, combining these methods with network panel data models \citep[see, for instance,][]{elhorst2014spatial,aquaro2015quasi,LESAGE20161}. 

Neglecting heterogeneities over time or the cross-section may conceal important transmission channels, for two reasons. First, there is evidence for structural breaks in the transmission of monetary policy shocks to macroeconomic and financial variables \citep{primiceri2005time,COGLEY2005262,paul2019time}. Several studies find that returns respond much stronger to surprise monetary policy shocks during tight credit market conditions, or during bear markets \citep[see][]{chen2007does,BASISTHA20082606,KUROV2010139,kontonikas2013stock}. It is unclear, however, if these differences originate from changes in the covariance structure across industries reflecting network dependency and higher-order effects, or whether they stem from direct responses in the conditional mean of conventional regressions (captured, for instance, via time-varying parameters, TVPs). While \citet{ozdagli2017monetary} assume constant parameters, they characterize the production network as non-linear and to exhibit cycles (see Section II of their paper). This motivates our approach of introducing time-varying network dependence alongside TVPs.

Second, pooling information across industries may conceal underlying structural relationships. And it potentially distorts the estimated importance of some industries in the disaggregate transmission of monetary policy shocks compared to others \citep[see][]{ehrmann2004taking,gorodnichenko2016sticky}. This is a crucial notion, considering that industries differ substantially in size and use vastly different production inputs. In a theoretical framework, \citet{pasten2019propagation,pasten2020price} show that differences in price rigidities originating from such heterogeneities are determinants how policy interventions are transmitted to the real economy.

To address heterogeneity over time and the cross-section, we develop a flexible Bayesian state-space model. Both the network dependence parameter and the regression coefficients are assumed to vary over time via imposing random walk state equations. The time-varying regression coefficients can be estimated by relying on a standard conditionally Gaussian state-space model using panel data for industry-level returns in the US. As a technical novelty, we moreover propose a sampling algorithm for the time-varying network dependence parameter. Our approach aims to shed light on the question whether network effects play a role in determining the overall time-varying impact of monetary policy shocks on stock returns.

From an empirical perspective, several findings are worth noting. First, we detect substantial differences over time and the cross-section. Our estimates indicate that the overall strength of network effects varies between $40$ percent and $80$ percent. Differences over time can be linked to periods of economic and financial uncertainty, often coinciding with tight credit market conditions and financial stress. Second, time-variation in network dependence translates to substantial differences in total effects of monetary policy on stock returns. In fact, we find that estimates in some periods are about two percent in response to a surprise one percentage point increase in the federal funds rate, while these effects can be as large as ten percent in others. Third, our results show substantial heterogeneity over the cross-section. We cluster industries by assessing the joint distribution of total and network effects econometrically, and obtain two main clusters. The clusters can roughly be described as classifying industries regarding their closeness to end-consumers in the production network. The closer an industry is to end-consumers, the smaller is the share attributed to network effects.

The rest of the paper is structured as follows. In Section \ref{sec:econometrics}, we set forth the model alongside the Bayesian prior setup and a sampling algorithm for inference. We apply the model in a study of the network effects of US monetary policy and discuss our findings in Section \ref{sec:application}. Section \ref{sec:conclusions} concludes.

\section{A time-varying network dependence panel model}\label{sec:econometrics}
We define the measurement equation for observation $i=1,\hdots,N$ as
\begin{equation}
y_{it} = \rho_t \sum_{j=1}^N w_{ijt} y_{jt} + \alpha_{it} + \bm{x}'_{it}\bm{\beta}_{it} + \epsilon_{it}, \quad \epsilon_{it}\sim\mathcal{N}\left(0,\sigma_{i}^2\right),\label{eq:model}
\end{equation}
where $y_{it}$ is the response variable at time $t=1,\hdots,T$. We include a time-varying intercept term $\alpha_{it}$, $K$ exogenous covariates in the $K\times1$-vector $\bm{x}_{it}$ with associated observation specific TVP vector $\bm{\beta}_{it}$ of size $K\times1$ and a Gaussian error term with zero mean and variance $\sigma_{i}^2$.

Information on the cross-sectional dependency structure is incorporated using weighted averages of the ``foreign'' quantities $y_{jt}~(j=1,\hdots,N)$ with time-varying weights $w_{ijt}$. These weights denote the elements of a pre-determined $N\times N$ weighting matrix $\bm{W}_t$ subject to the restrictions $w_{ijt}\geq0$ and $\sum_{j=1}^{N} w_{ijt} = 1$. Cross-sectional weights are commonly based on observables or simple ad hoc definitions, describing the network structure in a sensible way. We follow \citet{ozdagli2017monetary} and use a weights matrix capturing intermediate input shares across industries to model the US production network. The choice of this matrix is derived from a theoretical model of production with intermediate inputs and provides a precise structural interpretation. We explicitly allow for the network structure to change via $\bm{W}_t$ in our baseline specification to capture the varying relative importance of industries in the production network \citep[see also][]{carvalho2013great}.

We propose the scalar parameter $\rho_t$ to feature time-variation.\footnote{This feature is related to time-varying network structures \citep[see][]{asgharian2013spatial,billio2016impact}, assuming linkage matrices to evolve over time, but keeping the overall strength of network effects constant. By contrast, we introduce additional flexibility by assuming a time-varying network structure and dependence parameter. Our model can be considered as an extended Bayesian version of \citet{BLASQUES2016211} and \citet{catania2017dynamic} that features several technical novelties resulting in a more flexible specification.} The state equation for the network dependence parameter $\rho_t$ is a random walk process:
\begin{equation}
	\rho_{t}  = \rho_{t-1} + \varsigma\xi_{t}, \quad \xi_{t}\sim\mathcal{N}(0,1).\label{eq:tvrho}
\end{equation}

The covariance matrix of the reduced form errors for the stacked version of the model at time $t$ is given by the expression 
\begin{equation*}
(\bm{I}_N - \rho_t \bm{W}_t)^{-1}\bm{\Sigma}(\bm{I}_N - \rho_t \bm{W}_t)^{-1}{'},
\end{equation*}
with $\bm{\Sigma} = \text{diag}(\sigma_1^2,\hdots,\sigma_N^2)$. Econometrically, the parameter $\rho_t$ can thus be interpreted as a common factor, capturing a special form of stochastic volatility. $\bm{W}_t$ acts as a pre-determined matrix of factor loadings.\footnote{Although network multipliers (see next subsection) can also be estimated in unrestricted multivariate systems by decomposing the covariance matrix of the reduced form errors, identification of specific network connections and their interpretation is less straight-forward \citep{diebold2009measuring,bianchi2015modeling,billio2016interconnections}.} It relates to measures of dynamic connectedness \citep{diebold2009measuring,demirer2018estimating}, and studies capturing financial contagion and systemic risk \citep[see][]{forbes2002no,BLASQUES2016211}. The structural interpretation of the proposed $\bm{W}_t$ relates our study to investigations regarding network effects of aggregate demand shocks. Intuitively, since $\bm{W}_t$ solely captures time-varying relative input shares, $\rho_t$ governs time-varying cross-industry elasticities with respect to the exogenous variables \citep[see Section \ref{subsec:interpretation} in this paper, and Section III.A. in][for further details]{ozdagli2017monetary}.

Allowing for TVPs is straightforward by drawing from the vast literature on state-space models \citep[see][for a textbook overview]{kim1999state}. The regression coefficients are stacked in a $(K+1)\times1$-vector $\bm{\theta}_{it}=(\alpha_{it},\bm{\beta}_{it}')'$. We assume independent random walk state equations for industries $i=1,\hdots,N$:
\begin{equation*}
\bm{\theta}_{it} = \bm{\theta}_{it-1} + \bm{\eta}_{it}, \quad \bm{\eta}_{it} \sim \mathcal{N}(\bm{0},\bm{\Omega}_i).
\end{equation*}
Here, $\bm{\eta}_{it}$ is a zero-mean Gaussian error term and diagonal covariance matrix $\bm{\Omega}_i=\text{diag}(\omega_{i1},\hdots,\omega_{iK+1})$ of size $(K+1)\times(K+1)$. The state innovation variances in $\bm{\Omega}_i$ govern the degree of time-variation in the regression coefficients.

\subsection{Interpreting the model coefficients}\label{subsec:interpretation}
The approach to modeling network dependence pursued in this paper establishes a large system of simultaneous equations with specific parametric restrictions. Consequently, standard interpretations for linear regressions have to be adapted to account for the notion of cross-sectional dependencies. 

We follow \citet{LESAGE20161} and derive the impact matrix that contains the partial derivatives for all industries with respect to a change in the $k$th exogenous covariate $\bm{x}_{kt}=(x_{1kt},\hdots,x_{Nkt})$ of industry $i=1,\hdots,N$, $k=1,\hdots,K$, $t=1,\hdots,T$. Assuming time-varying network dependence and regression coefficients yields an impact matrix $\bm{S}_{kt}$:
\begin{equation*}
\frac{\partial \bm{y}_t}{\partial \bm{x}_{kt}} = \bm{S}_{kt} =
\begin{bmatrix}
\partial y_{1t}/\partial x_{1kt} & \partial y_{1t}/\partial x_{2kt} & \hdots & \partial y_{1t}/\partial x_{Nkt}\\
\partial y_{2t}/\partial x_{1kt} & \partial y_{2t}/\partial x_{2kt} & \hdots & \vdots \\
\vdots & \vdots & \ddots & \vdots \\
\partial y_{Nt}/\partial x_{1kt} & \hdots & \hdots & \partial y_{Nt}/\partial x_{Nkt}
\end{bmatrix}
 = (\bm{I}_N - \rho_t \bm{W}_t)^{-1} \bm{B}_{kt}.
\end{equation*}
Here, $\bm{B}_{kt} = \text{diag}(\beta_{1kt},\hdots,\beta_{Nkt})$ with $\beta_{ikt}$ referring to the $k$th coefficient of observation $i$ at time $t$, and the term $(\bm{I}_N - \rho_t \bm{W}_t)^{-1}$ is a network multiplier matrix governing the propagation of the shocks through the network structure. We define the following variants of impact effects:
\begin{itemize}[align=left]
	\item \textit{Direct effects per industry} are given by the main diagonal of $\bm{S}_{kt}$. This corresponds to the partial derivative of the response variable of industry $i$ with respect to the $k$th exogenous variable of the same industry adjusted for higher-order effects stemming from the network multiplier matrix. The \textit{average direct effect} is $1/N \times \text{tr}(\bm{S}_{kt})$, that is, the average of the main diagonal of the impact matrix $\bm{S}_{kt}$.
	\item The \textit{total effects per industry} can be calculated by $\bm{S}_{kt} \bm{\iota}_N$ (with $\bm{\iota}_N$ denoting an $N\times1$-vector of ones), reflecting the sum of all derivatives of the response variable in industry $i$ with respect to the $k$th explanatory variable of all other industries and itself. The \textit{average total effect} is defined as $1/N \times \bm{\iota}'_N \bm{S}_{kt} \bm{\iota}_N$.
	\item The \textit{average indirect effect} or network effect is the difference between the total and direct effects, and can also be computed per industry (\textit{indirect effects per industry}). This measure thus captures cross-industry partial derivatives on the off-diagonal positions in $\bm{S}_{kt}$. The share of \textit{network effects in percent} is calculated as indirect divided by total effects.
\end{itemize}

\subsection{Prior specification}\label{sec:priors}
We estimate the proposed model using Bayesian methods. This involves selecting suitable prior distributions for all parameters and combining them with the likelihood of the data. Conditional on a draw of the full history of this parameter $\{\rho_t\}_{t=1}^T$, inference for the other model parameters is standard. We choose the following prior distributions:
\begin{itemize}[align=left]
	\item To define the prior distribution on the time-varying regression coefficients, we consider the state-space model in its non-centered parameterization \citep[for details, see][]{FRUHWIRTHSCHNATTER201085}. Let $\sqrt{\bm{\Omega}_i} = \text{diag}(\sqrt{\omega_{i1}},\hdots,\sqrt{\omega_{iK+1}})$, then we split the coefficients into a constant and time-varying part: $\bm{\theta}_{it} = \bm{\theta}_{i0} + \sqrt{\bm{\Omega}_i}\bm{\tilde\theta}_{it}$. Using this transformation, $\bm{\tilde\theta}_{it}$ follows a random walk with standard normal shocks. For the prior on the initial state of the time-varying regression coefficients, we assume $\bm{\theta}_{i0} \sim \mathcal{N}(\bm{0},a\bm{V}_i)$ with $\bm{V}_i$ collecting the ordinary least squares variances on its main diagonal and $a=100$ determining the tightness the prior. This establishes a weakly informative variant of the g-prior \citep[see][]{Zellnergprior} for the time-invariant part of the coefficients. We use independent Gamma priors on the state innovation variances, which translates to a Gaussian prior on their square root \citep[see][]{FRUHWIRTHSCHNATTER201085}: $\sqrt{\bm{\Omega}_{i}}\sim\mathcal{N}(\bm{0},b\bm{V}_i)$. The tightness parameter $b$ is set to $0.1$, resulting in a comparatively tight prior that is required for regularizing the high-dimensional TVPs.
	\item For the initial state of the network dependence parameter $\rho_0$, we choose the prior $\rho_0\sim\mathcal{N}(\mu_0,\varsigma_0^2)$ with $\mu_0 = 0$ and $\varsigma_0^2=0.1$.
	\item On the state innovation variances of the network dependence parameter, we assume a mildly informative inverse Gamma prior, $\varsigma^2\sim\mathcal{G}^{-1}(c_\varsigma,d_\varsigma)$ with $c_\varsigma=3$ and $d_\varsigma=0.03$.
	\item The measurement equation error variances are assigned weakly informative independent inverse Gamma priors, $\sigma_i^{2}\sim\mathcal{G}^{-1}(c_\sigma,d_\sigma)$, with $c_\sigma=d_\sigma=0.01$.
\end{itemize}

\subsection{Estimating time-varying network dependence}\label{sec:posterior}
Combining the likelihood of the model with the proposed prior distributions yields a set of well-known conditional posterior distributions for most parameters that can be used for setting up a Markov Chain Monte Carlo (MCMC) sampling algorithm involving forward-filtering backward-sampling \citep[FFBS, see][]{doi:10.1093/biomet/81.3.541,doi:10.1111/j.1467-9892.1994.tb00184.x}. Most of the quantities involved are standard, and we discuss details in Appendix \ref{app:algorithm}. 

Producing draws for the full history of the time-varying network dependence parameter, however, is novel to the literature. In the following, we propose a sampling algorithm for the time-varying network dependence parameter. Due to the non-Gaussian setup, Kalman-filter based methods are inapplicable. Simulation from the posterior distribution can be carried out using a Metropolis-Hastings algorithm. We denote the current state of the respective quantity by $s-1$ and $s$ refers to a proposal from the candidate density. The procedure is similar to the algorithm proposed in the context of Bayesian stochastic volatility models in \citet{10.2307/1392151}. We rely on three proposal densities:
\begin{enumerate}[align=left]
	\item Since no initial value $\rho_{0}$ is available, we rely on \citet{10.2307/1392151} who show that this quantity can be obtained by drawing from a Gaussian distribution $\rho_0^{(s)}\sim\mathcal{N}(\bar{\mu}_{0},S_{0})$. The corresponding moments are $S_{0} = (\varsigma_0^2 \varsigma^2)/(\varsigma_0^2+\varsigma^2)$ and $\bar{\mu}_{0} = \varsigma_0^2(\mu_0/\varsigma_0^2 + \rho_1^{(s-1)}/\varsigma^2)$. The proposal at $t=1$ is then given by $\rho_1^{(s)}\sim\mathcal{N}(\bar{\mu}_1,S_1)$ where $\bar{\mu}_1 = (\rho_0^{(s)} + \rho_1^{(s-1)})/2$ and $S_1 = \varsigma^2/2$.
	\item For all points in time other than the first and last observation, a draw $\rho_{t}^{(s)}$ is generated from the proposal distribution given by $\rho_{t}^{(s)} \sim \mathcal{N}(\bar{\mu}_{t},S_{t})$, with $\bar{\mu}_{t} = (\rho_{t-1}^{(s)} + \rho_{t+1}^{(s-1)})/2$ and $S_{t} = \varsigma^2/2$.
	\item A similar problem arises for the final value at $t=T$, due to no $\rho_{T+1}$ being available. \citet{10.2307/1392151} suggest drawing from the modified candidate density $\rho_T^{(s)}\sim\mathcal{N}(\bar{\mu}_{T},S_{T})$ with $\bar{\mu}_{T} = \rho_{T-1}^{(s)}$ and $S_{T} = \varsigma^2$.
\end{enumerate}
For each point in time, we generate a proposal $\rho_t^{(s)}$ that can be used to calculate the acceptance probability of the Metropolis-Hastings algorithm. To simplify notation, we define $\tilde y_{it}(\rho_{t}^{(s)}) = \rho_t^{(s)} \sum_{j=1}^N w_{ij,t} y_{jt} \times \sigma_{i}^{-1}$ and $\tilde{\bm{y}}_t(\rho_{t}^{(s)}) = \left(\tilde y_{1t}(\rho_{t}^{(s)}),\hdots,\tilde y_{Nt}(\rho_{t}^{(s)})\right)'$ as the vector of network lags depending on the current value of $\rho_{t}^{(s)}$, with $\sigma_i^2$ referring to the error variance of industry $i$, and set $\tilde{\epsilon}_{it} = \left(y_{it} - \alpha_{it} - \bm{x}'_{it}\bm{\beta}_{it}\right) \times \sigma_{i}^{-1}$, where we again stack these quantities in $\bm{\tilde\epsilon}_t = (\tilde\epsilon_{1t},\hdots,\tilde\epsilon_{Nt})'$. Let 
\begin{equation*}
\mathcal{L}\left(\rho_{t}^{(s)}\right) = \det(\bm{I}_N - \rho_{t}^{(s)}\bm{W}_t) \times \exp\left\{-0.5\left(\bm{\tilde\epsilon}_t-\tilde{\bm{y}}_t(\rho_{t}^{(s)})\right)'\left(\bm{\tilde\epsilon}_t-\tilde{\bm{y}}_t(\rho_{t}^{(s)})\right)\right\},
\end{equation*}
then the acceptance probability $\zeta$ of the proposal $\rho_t^{(s)}$ implied by the likelihood is
\begin{equation*}
\zeta=\min\left(
	\frac{\mathcal{L}\left(\rho_{t}^{(s)}\right)}{\mathcal{L}\left(\rho_{t}^{(s-1)}\right)},1\right).
\end{equation*}
The candidate draw $\rho_t^{(s)}$ is accepted with probability $\zeta$. Otherwise, we retain the previous draw $\rho_t^{(s-1)}$. After obtaining the full history for $\rho_{t}$, we simulate the variance $\varsigma^2$ using standard posterior moments for the error variance in Bayesian linear regression models. 

\section{Network effects of US monetary policy}\label{sec:application}
\subsection{Data and model specification}\label{sec:datamodspec}
In this subsection we describe that dataset. We first provide information on the exogenous monetary policy shocks. This discussion is followed by our classification of industries and the construction of the cross-sectional linkages.

\subsubsection{Measuring monetary policy shocks}
As exogenous measure of the monetary policy shocks, we rely on high-frequency changes in Federal funds futures. The predetermined nature of monetary policy announcement dates (eight regular FOMC meetings per year, with press releases communicating policy decisions typically around 14:15 Eastern time) allows for extracting the surprise component of the monetary policy action. We use high-frequency data on forward-looking financial instruments in a tight window of $\Delta t = \tau(1) + \tau(2) = 30$ minutes around the press release. In particular, we define monetary policy shocks $v_t$ as:
\begin{equation*}
v_t = \frac{D}{D-t}\left(\text{FF}_{t+\tau(2)}-\text{FF}_{t-\tau(1)}\right).
\end{equation*}
$\text{FF}_{t+\tau(2)}$ is the rate implied by federal funds futures after the announcement at time $t$, while $\text{FF}_{t-\tau(1)}$ denotes the same rate before the FOMC announcement. $D$ is the number of days in the month, which is needed for adjusting for the fact that the federal funds futures settle on the average effective overnight federal funds rate. The tight window around the announcement defined by $\tau(1)=10$ minutes and $\tau(2)=20$ minutes reduces the risk of other events than monetary policy decisions affecting futures prices and provides support for the claim of exogeneity \citep[see also][]{bernanke2005explains,gurkaynak2005actions}.

We focus on scheduled Federal Open Market Committee (FOMC) meetings and exclude emergency meetings to reduce the risk of biasing our estimates with confounding signaling effects \citep[see, for instance,][]{nakamura2018high,jarocinski2018deconstructing}. Our information set includes data between February 1994 and December 2008, that is, $T=120$. The sample starts in 1994 because the Federal Reserve (Fed) changed its communication strategy at this time and tick-by-tick stock market data is not available prior to 1993. It ends in 2008 to exclude the period when the Fed started its various unconventional monetary policy measures when approaching the zero lower bound.

The exogenous vector $\bm{x}_{it}$ in \autoref{eq:model} features the scalar shock $v_t$ that is common to all $i$, while $\beta_{it}$ is the associated time-varying observation-specific parameter capturing the sensitivity of industry $i$ to the monetary policy shock at time $t$. Moreover, we include an industry-specific constant $\alpha_{it}$.

\subsubsection{Industry-level event returns}
The industries are selected based on the availability of input-output (IO)-tables published by the \textit{Bureau of Economic Analysis} (BEA) and the \textit{United States Department of Commerce}. These tables are needed to calculate the cross-sectional linkages in $\bm{W}_t$. They are published every five years, and we utilize their 1992, 1997 and 2002 versions.

We aggregate industries at the four-digit IO aggregation level, which can be mapped to the Standard Industrial Classification (SIC) and North American Industry Classification System (NAICS). The event returns for industry $i$ used as dependent variables $y_{it}$ are constructed based on returns for all common stocks trading on the NYSE, Amex or Nasdaq around press releases by the FOMC, weighted by the corresponding market capitalization at the end of the previous trading day for industries $i=1,\hdots,N$. The dependent variable is defined as the difference between the last trade observation before, and the first observation after the event window. Note that we exclude industries with less than three firms to ensure diversified industry returns and limit the risk of outliers affecting our results.

Industry classifications change between 1992 and subsequent IO-table publications. For our main results in Section \ref{sec:results}, we rely on the codes in use from 1997 onwards. The panel framework requires consistent availability of event returns over time. Following \citet{ozdagli2017monetary}, we exclude zero event returns, which results in $N=58$ industries in our baseline specification. Details on the industries are provided in Appendix \ref{app:data}. For the robustness checks provided in Appendix \ref{app:robustness}, we also present estimates using a time-invariant weighting matrices, resulting in different numbers of available non-zero industry-returns due to differences in the aggregation scheme governed by the IO-tables.

\subsubsection{Cross-sectional dependency}
To establish the cross-sectional dependency structure via the weighting matrix $\bm{W}_t$ we use IO-tables capturing dollar trade flows between industries. The BEA provides so-called ``make'' (denoted by an industry-by-commodity matrix $\bm{W}^{\text{(make)}}_{t}$ of size $N\times C$ with elements $w_{ict}^{\text{(make)}}$, the production of goods by industries) and ``use'' tables (denoted by a commodity-by-industry matrix $\bm{W}^{\text{(use)}}_{t}$ of size $C\times N$ with elements $w_{cjt}^{\text{(use)}}$, the uses of commodities by intermediate and final users). 

Following \citet{ozdagli2017monetary}, we define the market shares $\bm{W}^{\text{(share)}}_{t}$ of the production industries as
\begin{equation*}
w_{ict}^{\text{(share)}} = w_{ict}^{\text{(make)}} / \sum_{i=1}^{N} w_{ict}^{\text{(make)}}.
\end{equation*}
The share and use tables are used to calculate the amount of dollars industry $j$ sells to industry $i$, denoted by the $N\times N$-matrix $\bm{W}^{\text{(rev)}}_{t}$:
\begin{equation*}
\bm{W}^{\text{(rev)}}_{t} = \bm{W}^{\text{(share)}}_{t}\bm{W}^{\text{(use)}}_{t}.
\end{equation*}
The final step uses this matrix to derive the percentage of industry $i$ inputs purchased from industry $j$, which defines the elements of the weight matrix $\bm{W}_t$ introduced in Section \ref{sec:econometrics}:
\begin{equation*}
w_{ijt} = w_{ijt}^{\text{(rev)}} / \sum_{c=1}^C w_{cjt}^{\text{(use)}}.
\end{equation*}

In our baseline model, we allow for time-variation in $\bm{W}_t$. We achieve this by using the consistently available coding of industries starting 1997, using the 1997 IO-tables from 1994 to the last FOMC announcement in 2001, and rely on the 2002 IO-tables from this point onwards. This specifications allows for changes in the strength of overall network dependence, while addressing changes in the overall structure of industry relations via the weights matrix.

\subsection{Empirical results}\label{sec:results}
In a first step, we compare the results estimated with our proposed model to a set of related specifications from the established literature. For the models featuring heterogeneous coefficients, we take the arithmetic mean over all industries and over time per iteration of the algorithm and report the resulting posterior percentiles (the posterior median, and the bounds marking the $99$ percent posterior credible set). This provides a measure of the average impact of monetary policy shocks on heterogeneous industry returns. 

The different specifications are summarized in Table \ref{tab:models}: ``\textit{Data}'' indicates whether the model was estimated using aggregate (S\&P 500) or granular industry-specific data (Industries). The aggregate S\&P 500 returns in 30-minute windows around FOMC announcement dates are taken from \citet{gorodnichenko2016sticky}, and the exercise corresponds roughly to \citet{bernanke2005explains} and \citet{gurkaynak2005actions}. The industry-level data is constructed as discussed in Section \ref{sec:datamodspec}. 

``\textit{Heterogeneity}'' marks which coefficients allow for heterogeneity. Relevant cases are pooled specifications over time and the cross-section (--), implying that we rule out time-variation in the coefficients and set $\bm{\theta}_1=\hdots=\bm{\theta}_N$ and $\sigma^2_1,=\hdots=\sigma^2_N$. Specifications marked with an $i$ indicate that we allow for industry-specific coefficients $\bm{\theta}_i$ and $\sigma^2_i$, but surpress time-variation in the regression coefficients. Those marked with $t$ refer to time-varying regression coefficients (relevant only for the aggregate data), while $i,t$ refers to all parameters being estimated freely across industries and over time. All of these specifications are nested in our proposed model. 

``\textit{Network}'' refers to the specification of the network dependence parameter: -- means no network dependence, $\rho$ marks constant network dependence, and $\rho_t$ refers to the time-varying network dependence model proposed in this paper. The weights matrix $\bm{W}_t$ features time variation and is described in detail in Section \ref{sec:datamodspec}. 

``\textit{References}'' provides an overview of references to similar specifications in the literature referred to in Section \ref{sec:introduction}. Note that \citet{chen2007does} and \citet{BASISTHA20082606}, referenced in the context of time-varying parameter specifications, rely on a different specification of the TVPs using regime-switching models. By contrast, we allow for gradual changes in the network dependence parameter and the regression coefficients.

\begin{table*}[t]
\caption{Model specifications.}\vspace*{-1.5em}
\begin{center}
\begin{footnotesize}
\begin{threeparttable}
\begin{tabular*}{\textwidth}{@{\extracolsep{\fill}} llcll}
\toprule
Model & Data & Heterogeneity & Network & References\\
\midrule
  A1 & S\&P 500 & -- & -- & \citet{gurkaynak2005actions} \\ 
  A2 & S\&P 500 & $t$ & -- & \citet{chen2007does}\\ 
  \midrule
  B1 & Industries & -- & -- & \citet{bernanke2005explains} \\ 
  B2 & Industries & $i$ & -- & \citet{ehrmann2004taking} \\ 
  B3 & Industries & -- & $\rho$ & \citet{ozdagli2017monetary} \\ 
  B4 & Industries & $i$ & $\rho$ & --- '' ---\\ 
  \midrule
  C1 & Industries & -- & $\rho_t$ & \\ 
  C2 & Industries & $i$ & $\rho_t$ & \\ 
  C3 & Industries & $i,t$ & -- & \citet{BASISTHA20082606} \\ 
  C4 & Industries & $i,t$ & $\rho$ & \\ 
  C5 & Industries & $i,t$ & $\rho_t$ & \\ 
\bottomrule
\end{tabular*}
\begin{tablenotes}[para,flushleft]
\scriptsize{\textit{Notes}: ``Data'' indicates whether the model was estimated using aggregate (S\&P 500) or granular industry-specific (Industries) data. ``Heterogeneity'' marks whether we pool estimates over time and the cross-section (--), allow for variation over the cross-section ($i$), over time ($t$), or the cross-section and over time ($i,t$). ``Network'' refers to the specification of the network dependence parameter: -- means no network dependence, $\rho$ marks constant network dependence, and $\rho_t$ refers to time-varying network dependence model proposed in this paper. ``References'' provides references to similar specifications in the established literature.}
\end{tablenotes}
\end{threeparttable}
\end{footnotesize}
\end{center}
\label{tab:models}
\end{table*}

The results across the different model types are displayed in Table \ref{tab:results}. For the cases where there is no network dependence or where we rely on aggregate data, the regression coefficient associated with the monetary policy shocks corresponds to the total effect (no spillovers). Negative values for total impacts imply stock market responses in line with standard economic theory. Monetary tightening induces a reduction of future expected dividends, and by basic asset pricing theory, higher interest rates increase the discout rate of future dividends, resulting in stock market declines. Robustness checks showing very similar results for different specifications of the weights matrix or industry aggregations, alongside a split-sample analysis, are provided in Appendix \ref{app:robustness}.

We start by comparing the disaggregate, industry-based estimates with those obtained from regressing aggregate S\&P 500 returns around announcement dates on the monetary policy shocks displayed in the first row of Table \ref{tab:results}. For this purpose, we replicate the setup in \citet{gurkaynak2005actions}, who rely on data from January 1990 to December 2004, using our updated dataset from February 1994 to December 2008. At this point, we note that our estimates of the total effects are rather similar for point estimates across all different specifications (with minor differences in posterior credible sets), indicating that our proposed model produces reasonable results in line with the established literature.

\begin{table*}[t]
\caption{Estimated impacts of monetary policy on stock returns across industries.}\vspace*{-1.5em}
\begin{center}
\begin{scriptsize}
\begin{threeparttable}
\begin{tabular*}{\textwidth}{@{\extracolsep{\fill}} lcccccccc}
\toprule
               & \multicolumn{4}{c}{\textbf{Parameters}} & \multicolumn{4}{c}{\textbf{Impacts}}\\
               \cmidrule(lr){2-5}\cmidrule(lr){6-9}
   & $\alpha$ & $\beta$ & $\sigma^2$ & $\rho$ & Indirect & Direct & Total & Netw. (\%) \\ 
\midrule
A1 & -0.13 & -3.11 & 0.23 &  &  & -3.11 & -3.11 &  \\ 
   & (-0.24, -0.02) & (-4.90, -1.17) & (0.17, 0.33) &  &  & (-4.90, -1.17) & (-4.90, -1.17) &  \\ 
  A2 & -0.15 & -3.49 & 0.19 &  &  & -3.49 & -3.49 &  \\ 
   & (-0.25, -0.04) & (-5.45, -1.58) & (0.14, 0.27) &  &  & (-5.45, -1.58) & (-5.45, -1.58) &  \\ 
   \midrule
  B1 & -0.12 & -3.19 & 0.15 &  &  & -3.19 & -3.19 &  \\ 
   & (-0.14, -0.11) & (-3.49, -2.89) & (0.14, 0.15) &  &  & (-3.49, -2.89) & (-3.49, -2.89) &  \\ 
  B2 & -0.12 & -3.01 & 0.34 &  &  & -3.01 & -3.01 &  \\ 
   & (-0.14, -0.11) & (-3.30, -2.75) & (0.32, 0.36) &  &  & (-3.30, -2.75) & (-3.30, -2.75) &  \\ 
  B3 & -0.04 & -1.05 & 0.15 & 0.67 & -1.78 & -1.40 & -3.19 & 56.0 \\ 
   & (-0.06, -0.03) & (-1.27, -0.86) & (0.14, 0.15) & (0.65, 0.70) & (-2.16, -1.49) & (-1.67, -1.15) & (-3.85, -2.64) & (53.6, 58.6) \\ 
  B4 & -0.02 & -0.49 & 0.13 & 0.84 & -2.19 & -0.79 & -2.98 & 73.5 \\ 
   & (-0.03, -0.01) & (-0.66, -0.29) & (0.12, 0.14) & (0.82, 0.86) & (-2.97, -1.42) & (-1.04, -0.53) & (-3.97, -1.99) & (70.6, 76.2) \\ 
   \midrule
  C1 & -0.05 & -1.08 & 0.15 & 0.52 & -1.43 & -1.34 & -2.78 & 51.8 \\ 
   & (-0.07, -0.04) & (-1.32, -0.84) & (0.14, 0.16) & (0.49, 0.55) & (-1.87, -1.13) & (-1.63, -1.06) & (-3.43, -2.22) & (48.5, 56.1) \\ 
  C2 & -0.03 & -0.61 & 0.13 & 0.72 & -2.04 & -0.88 & -2.93 & 69.7 \\ 
   & (-0.04, -0.02) & (-0.82, -0.43) & (0.12, 0.14) & (0.70, 0.75) & (-2.92, -1.44) & (-1.16, -0.66) & (-4.01, -2.05) & (66.3, 74.8) \\ 
  C3 & -0.12 & -3.05 & 0.34 &  &  & -3.05 & -3.05 &  \\ 
   & (-0.14, -0.11) & (-3.36, -2.78) & (0.32, 0.36) &  &  & (-3.36, -2.78) & (-3.36, -2.78) &  \\ 
  C4 & -0.02 & -0.49 & 0.12 & 0.85 & -2.54 & -0.84 & -3.37 & 75.2 \\ 
   & (-0.03, -0.01) & (-0.67, -0.29) & (0.12, 0.14) & (0.83, 0.86) & (-3.41, -1.80) & (-1.09, -0.59) & (-4.52, -2.40) & (72.9, 77.5) \\ 
  C5 & -0.03 & -0.63 & 0.13 & 0.74 & -2.41 & -0.96 & -3.36 & 71.3 \\ 
   & (-0.04, -0.02) & (-0.82, -0.42) & (0.12, 0.14) & (0.72, 0.77) & (-3.64, -1.60) & (-1.22, -0.69) & (-4.73, -2.28) & (67.5, 78.1) \\ 
\bottomrule
\end{tabular*}
\begin{tablenotes}[para,flushleft]
\scriptsize{\textit{Notes}: For model specifications see Table \ref{tab:models}. For those featuring heterogeneous coefficients over the cross-section $i$ or over time $t$, we take the arithmetic mean over all industries and over time per iteration of the algorithm and report the resulting posterior percentiles (the posterior median, and the bounds in parentheses marking the $99$ percent posterior credible set). Impact effects are defined as in Section \ref{subsec:interpretation}.}
\end{tablenotes}
\end{threeparttable}
\end{scriptsize}
\end{center}
\label{tab:results}
\end{table*}

Accounting for posterior uncertainty and the different sample period, our estimates for the aggregate Model A1 corroborate the findings in \citet{gurkaynak2005actions}. A surprise one percentage point increase in the federal funds rate translates to a decline in stock market returns of about $3.1$ percent. Considering a hypothetical positive $25$ basis point (bp) shock to the federal funds rate -- the usual magnitude of Fed policy adjustments for the considered period -- yields a decline of the S\&P 500 index between $1.2$ and $0.3$ percent. These effects are in line with \citet{gurkaynak2005actions} and \citet{bernanke2005explains}, and also mirror those of \citet{ozdagli2017monetary} in the context of an identical replication exercise for our sample period. Allowing for time-variation in the coefficient measuring the sensitivity of S\&P 500 returns to monetary policy shocks (Model A2) and aggregating the response over time ex post yields marginally larger point estimates with a slightly inflated posterior credible interval. We discuss time-varying dynamics below, but note that effect sizes differ strongly over time, a finding in line with \citet{chen2007does}.

The results for models estimated with industry-level data, disregarding time-variation in the regression coefficients or the network dependence parameter for the moment, are summarized in rows two to six (labeled Model B1 to B4) in Table \ref{tab:results}. Starting with Models B1 and B2, abstracting from higher-order effects captured by network dependence models, we find point estimates to be similar to those obtained from estimating the model using aggregate data. It is worth mentioning that the posterior credible sets are much narrower. We provide a detailed discussion of cross-sectional heterogeneity below, but note that our estimates corroborate the notion of asymmetric effects of monetary policy shocks on industry-returns, as suggested by \citet{ehrmann2004taking}.

Crucial benchmarks are Model B3 and B4, which are the main specifications in \citet{ozdagli2017monetary}. Recall that our proposed specifications feature the time-varying weights matrix $\bm{W}_t$ and are estimated using the balanced panel of $N=58$ industries. Compared to the original paper, the estimates are remarkably robust to this different sample in terms of total effect sizes. However, our estimates for the parameter $\rho_t$ are appreciably lower. While \citet{ozdagli2017monetary} estimate the network dependence parameter for the homogeneous coefficient specification (Model B3) to be around $0.87$, ours lies in the credible set between $0.65$ and $0.7$. Turning to Model B4 featuring idiosyncratic regression coefficients and variances, our results are almost identical to those presented in Table 2, column 5 in \citet{ozdagli2017monetary}, the the corresponding specification. Calculating relative network effects, this implies that roughly $74$ percent of the overall market response can be explained by higher-order effects.

Specifications featuring TVPs or a time-varying network dependence parameter are shown in the bottom panel of Tabel \ref{tab:results} (labeled Model C1 to C5). Starting with Model C1, ruling out TVPs and pooling over the cross-section but allowing for a time-varying $\rho_t$, we find total impacts to be slightly lower than in all others. The estimates for network effects in percent are comparable to Model B3. Relaxing the assumtion of homogeneity over the cross-section increases the share attributed to network effects substantially. Estimated impacts and network effects are similar to those in Model B4, the main specification of \citet{ozdagli2017monetary}. For Model C3, where we neglect higher-order effects, we find average total impacts of $-3.05$ percent in response to a surprise one percentage point increase in the federal funds rate. These estimates are smaller in size compared to \citet{BASISTHA20082606}. Note, however, that we rely on a different sampling period, and their estimates are solely based on the aggregate S\&P 500 index. Moreover, rather than relying on a regime-switching model, we allow for gradually evolving coefficients and observe substantial variation in the effects over time.

Model C4 and C5 reflect variants of our main specification. Several findings are worth noting. First, we obtain significantly larger estimates for the network dependence parameter if we rule out time-varying network dependence. This translates to a slightly higher share of the total effects attributed to higher-order network effects of about $75$ percent. Relaxing the assumption of constant regression coefficients slightly increases (decreases) our estimates for direct effects (indirect effects). This dynamic yields an estimate for the network effects between $67.5$ to $71.3$ percent, leaving the total effects roughly unchanged. Interestingly, our estimates for total effects are comparable to Model A2 using aggregate data, albeit with narrower credible sets.

Summing up, we observe small differences across the model specifications. However, all of them are in line with the established literature and our proposed modeling approach appears to deliver plausible results. In the following, we illuminate driving factors of these differences based on cross-sectional heterogeneities, time-variation in  regression coefficients and the network dependence parameter.

\subsubsection{Time-varying effects of monetary policy shocks on stock returns}
\begin{figure}[t]
\includegraphics[width=\textwidth]{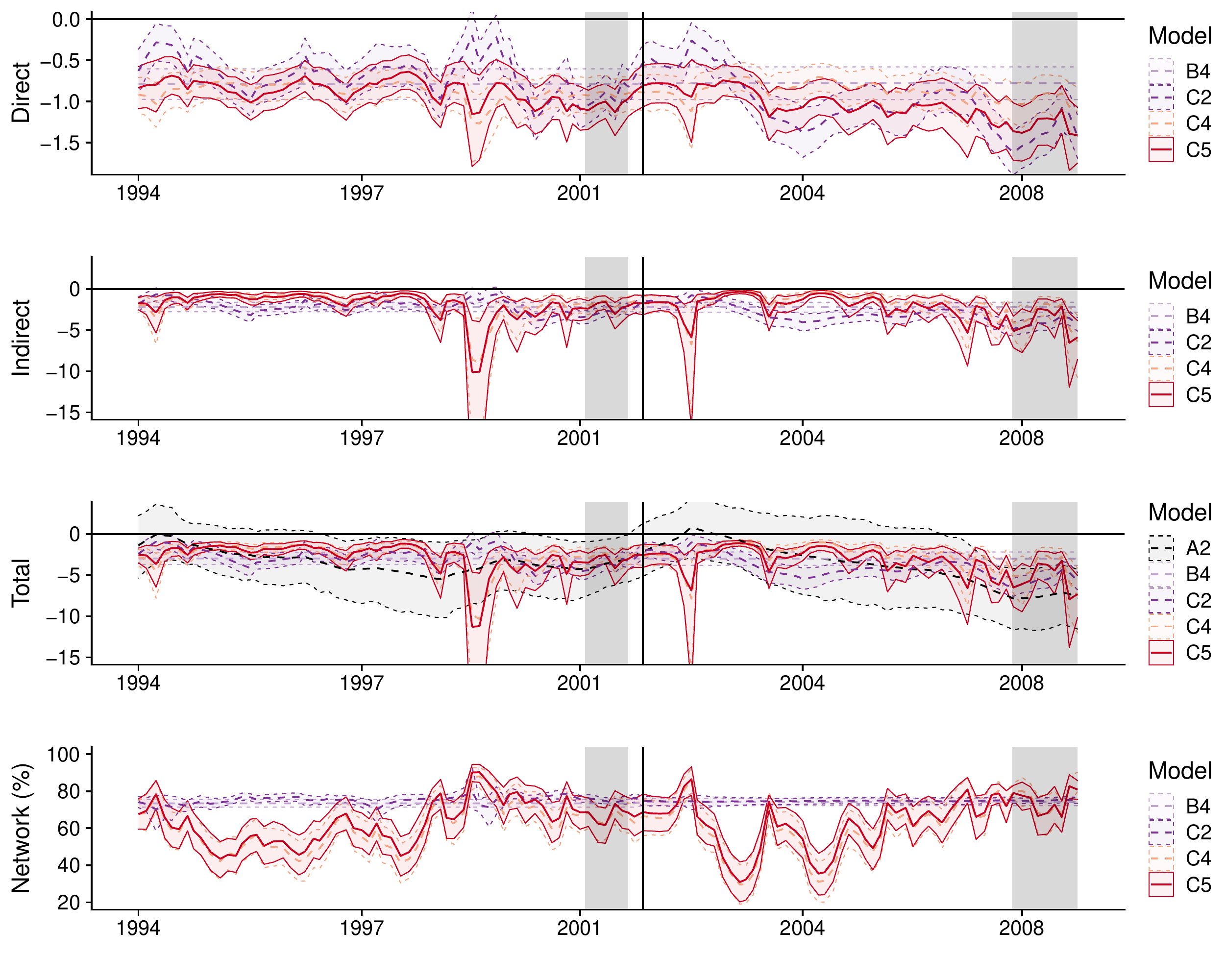}
\caption{Impact effects over time across different model specifications.}\label{fig:tvp-fx}\vspace*{-0.3cm}
\caption*{\footnotesize\textit{Note}: Details on the impact measures are described in Section \ref{subsec:interpretation}. For model specifications see Table \ref{tab:models}. Solid lines and shaded areas depict the $99$ percent posterior credible set and the posterior median. The grey shaded area marks recessions dated by the NBER Business Cycle Dating Committee. The vertical black solid line indicates the January 30th 2002 policy meeting, where the weights matrix changes.}
\end{figure}

In this section, we investigate average impact effects over time. Direct, indirect, total and network effects in percent are displayed in Figure \ref{fig:tvp-fx}. We focus on Models B4 (constant parameter benchmark model), C2 and C4, and compare them to our main specification C5. As an aggregate benchmark, we also include Model A2. The models are selected based on illuminating differences over time arising from introducing different types of heterogeneities.

Before turning to explanations of why impacts change over time, we provide a description of the estimated impact effects. Several findings are worth noting. Direct effects mostly exhibit a smooth path, albeit with several high-frequency spikes. Differences across model specifications featuring TVPs appear especially in 1999 and between 2002/2003. In particular, Model C2 estimates much smaller effects in absolute value alongside movements in the opposite direction when compared to C4 and C5. It is worth mentioning that indirect effects for C2 are extremely smooth over time (and look similar to direct effects, given the constant specification of $\rho_t$), while the models featuring a time-varying network parameter exhibit numerous high-frequency spikes. Comparing Models C4 and C5 in detail and assessing the effect of allowing for time-varying regression coefficients, we find that differences are muted. We estimate slightly larger direct effects in absolute value for C4, but the dynamic evolution of the impact measures is rather similar.

One of the main questions this paper aims to address is how total effects of monetary policy on stock market returns evolve over time. The third panel in Figure \ref{fig:tvp-fx} shows these effects for several models estimated using industry-level data, and also plots Model A2 which is based on the aggregate S\&P 500 index. With aggregate data, the credible sets are inflated and include zero for a substantial part of the sample. Time-variation in the estimates is occuring at a rather low frequency, similar to C2. The overall dynamic evolution is comparable to models C4 and C5, although we observe differences in 1999 and 2002/2003. These differences can be explained by the fact that the time-varying network dependence model allows for shifts in the covariance structure across industries. We refer to the discussion of interpreting $\rho_t$ as a common factor capturing a form of stochastic volatility. Trends towards larger effects at the end of the sample are clearly visible.

Part of the total impact can be explained by higher-order network effects, which are shown in the bottom panel in percent. We observe that network effects for Models C4 and C5 are approximately the same. Similarly, C2 and B4 are rather similar, and correspond to the average of C4 and C5 over time (about $80$ percent). Interestingly, we observe substantial variation in the strength of network effects over time. Between 1994 and 1998, about $50$ percent of the total impact can be explained by network effects. After a period of elevated network effects and several higher-frequency movements exceeding $80$ percent, we observe the posterior median to drop to about $40$ percent. Twoards the end of the sample, we estimate a persistently high importance of network effects of around $80$ percent.

\begin{figure}[t]
\includegraphics[width=\textwidth]{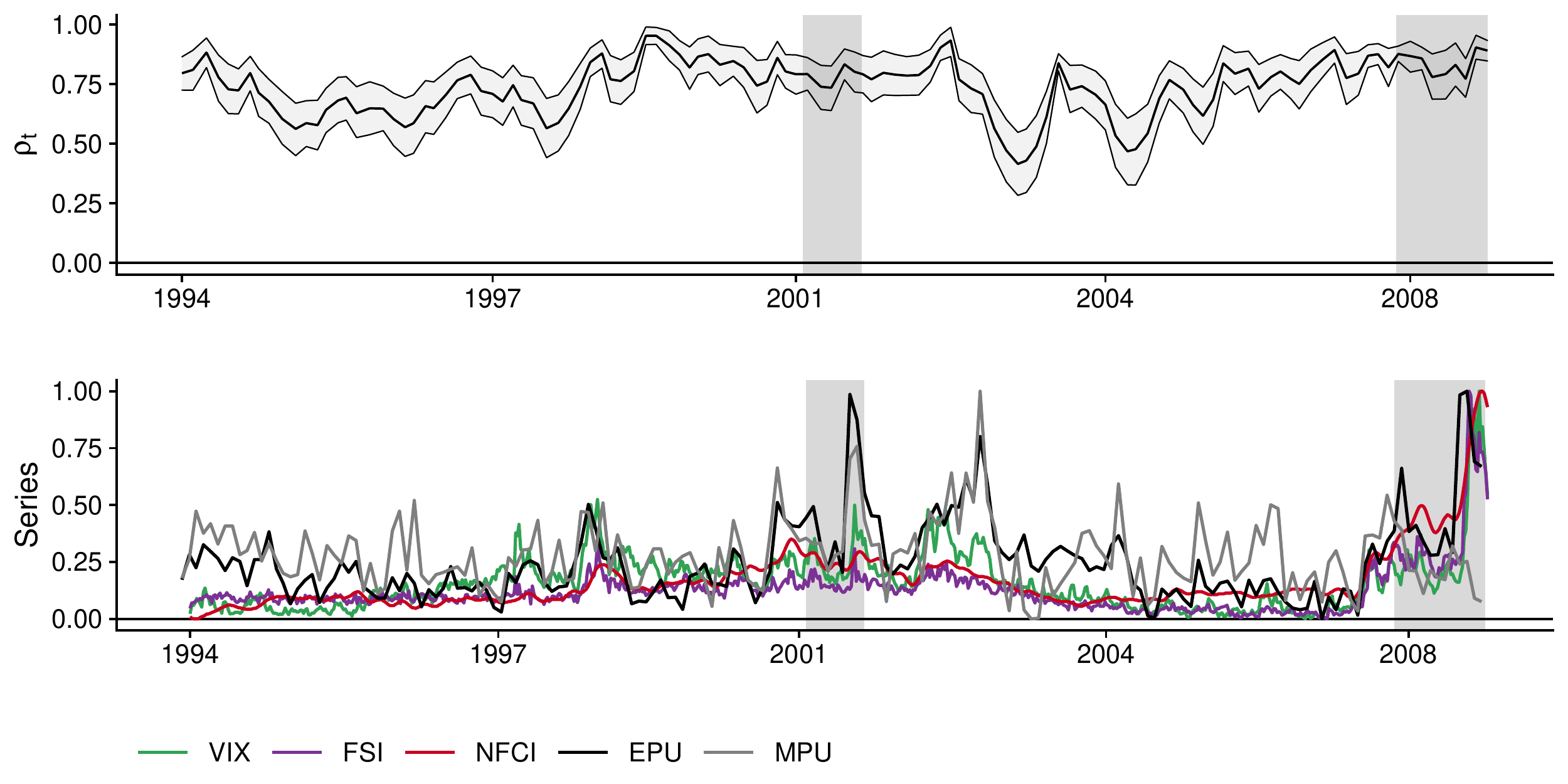}
\caption{Explaining time-variation in the network dependence parameter.}\label{fig:netw-ex}\vspace*{-0.3cm}
\caption*{\footnotesize\textit{Note}: Posterior median of the network dependence parameter $\rho_t$ alongside the $99$ percent posterior credible set. ``VIX'' is the Chicago Board Options Exchange Volatility Index, ``FSI'' is the St. Louis Fed Financial Stress Index, ``NFCI'' is the Chicago Fed National Financial Conditions Credit Subindex and ``EPU'' is the three-component economic policy uncertainty index developed by \citet{doi:10.1093/qje/qjw024}. These series are normalized such that they lie in the unit interval to be comparable in scale. For values $x$, with $\min$ and $\max$ referring to the minimum and maximum values, the normalization is $[x - \min(x)]/[\max(x)-\min(x)]$. The grey shaded areas marks recessions dated by the NBER Business Cycle Dating Committee.}
\end{figure}

As next step, we investigate the time-varying dependence parameter $\rho_t$ in the upper panel of Figure \ref{fig:netw-ex}. Comparing this time-series to indirect, total and network effects in Figure \ref{fig:tvp-fx}, high-frequency movements are clearly driven by the network dependence parameter. Recall that the parameter $\rho_t$ can be interpreted as a common factor scaling the covariance matrix of the reduced form errors, and thus captures a special form of stochastic volatility. The lower panel of Figure \ref{fig:netw-ex} collects several series that we link to the observed dynamics in higher-order network effects of monetary policy to explain the time-variation.

The related literature provides several potential explanations for time-variation in the transmission of monetary policy interventions. They include differences across investor sentiments over stock market regimes (bull and bear markets), credit conditions and financial stress, but also financial and economic uncertainty \citep[see][]{chen2007does,KUROV2010139,kontonikas2013stock,baker2019policy,husted2019monetary}. 

We focus on five series of interest that reflect such conditions. They are obtained from the FRED database maintained by the \textit{Federal Reserve Bank of St. Louis}, and normalized to lie in the unit interval to make them commensurable in scale. We include the Chicago Board Options Exchange Volatility Index (VIX), which captures the stock market's expectation of volatility based on S\&P 500 index options. The VIX captures overall financial market uncertainty. Moreover, we investigate the St. Louis Fed Financial Stress Index (FSI) and the Chicago Fed National Financial Conditions Credit Subindex (NFCI). These indices serve as measurements for financial stress and the tightness of credit market conditions. As a broader measure of uncertainty, we refer to the economic policy uncertainty (EPU) index developed by \citet{doi:10.1093/qje/qjw024}, accompanied by a measure of monetary policy uncertainty \citep[MPU,][]{husted2019monetary}.\footnote{The economic policy and monetary policy uncertainty indices \citep[see][]{doi:10.1093/qje/qjw024,husted2019monetary} are obtained from \href{https://www.policyuncertainty.com}{policyuncertainty.com}.}

It is worth mentioning that all series exhibit a substantial degree of comovement, with EPU and MPU showing several differences particularly between 2002 and 2005. Table \ref{tab:cormat} shows pairwise correlations. If publication frequencies are higher than monthly, we aggregate them at a monthly frequency using the arithmetic mean and match them with the FOMC meeting dates. The network dependence parameter exhibits the highest correlation with NFCI, followed by FSI and the VIX. This points towards the importance of financial uncertainty increasing higher-order demand effects of monetary policy, alongside tight credit market conditions.

\begin{table*}[t]
\caption{Correlation matrix of the network dependence parameter with various indices.}\vspace*{-1.5em}
\begin{center}
\begin{footnotesize}
\begin{threeparttable}
\begin{tabular*}{\textwidth}{@{\extracolsep{\fill}} llllll}
  \toprule
 & $\rho_t$ & NFCI & FSI & VIX & EPU \\ 
  \midrule
  NFCI & 0.533$^{***}$ &     &     &     &     \\ 
  FSI & 0.447$^{***}$ & 0.720$^{***}$ &     &     &     \\ 
  VIX & 0.429$^{***}$ & 0.623$^{***}$ & 0.845$^{***}$ &     &     \\ 
  EPU & 0.205$^{*}$ & 0.491$^{***}$ & 0.611$^{***}$ & 0.525$^{***}$ &     \\ 
  MPU & 0.255$^{**}$ & 0.208 & 0.314 & 0.290$^{**}$ & 0.501$^{***}$ \\ 
   \midrule
\end{tabular*}
\begin{tablenotes}[para,flushleft]
\scriptsize{\textit{Notes}: ``VIX'' is the Chicago Board Options Exchange Volatility Index, ``FSI'' is the St. Louis Fed Financial Stress Index, ``NFCI'' is the Chicago Fed National Financial Conditions Credit Subindex, ``EPU'' is the three-component economic policy uncertainty index developed by \citet{doi:10.1093/qje/qjw024} and ``MPU'' the monetary policy uncertainty index of \citet{husted2019monetary}. If publication frequencies are higher than monthly, we aggregate them at a monthly frequency using the arithmetic mean and match them with FOMC meeting dates. Asterisks indicate p-values: $0.001$ ($^{\ast\ast\ast}$), $0.01$ ($^{\ast\ast}$), $0.05$ ($^{\ast}$).}
\end{tablenotes}
\end{threeparttable}
\end{footnotesize}
\end{center}
\label{tab:cormat}
\end{table*}

The first substantial peak occurs during the Asian financial crisis in 1997, followed by the Russian crisis and the related collapse of the hedge-fund long-term capital management in late 1998. During these periods, all measures inidcate elevated levels, pointing towards these events increasing US stock market volatility, uncertainty, and financial stress. The second major peak occurs in the context of the burst of the dot-com bubble in 2000. From this point on, network dependence is persistently high, with minor high-frequency movements during the 9/11 terrorist attacks and the outbreak of Gulf War II. The latter is mainly observable in the EPU and MPU indices, pointing towards increased demand effects of monetary policy measures during periods of high economic uncertainty. Significant drops are observable in early 2003 and mid 2004, periods where EPU and MPU show large decreases. We detect persistently increasing high network dependence up to the collapse of Lehman Brothers in late 2008.  

Our findings corroborate those of the earlier literature that time-variation in stock market responses to monetary policy shocks are related to economic and financial uncertainty, investor sentiment in bull and bear markets and financial stress and credit market conditions.

\subsubsection{Assessing heterogeneity and clustering of industries}
In this section, we shed light on industry-specific effects over time. As a first step, we abstract from the time dimension and assess clusterings of industries based on average values over the full sample period. The methods proposed in our paper do not allow for clustering the impacts in a unified econometric approach. This is due to non-linearities in the conditional mean of the model, and because the effects of interest are non-linear functions depending the reduced form parameters (see Section \ref{subsec:interpretation}). 

As a solution, we rely on $k$-means clustering of industries using the joint distribution of total and network effects based on each individual draw from the posterior. We choose total and network effects for assessing clusters based on arguments of structural differences arising from how close the respective industries are to end-consumers, provided in \citet{ozdagli2017monetary}. 

Our analysis requires the number of clusters $k$ to be chosen a priori. A common way to choose $k$ is to rely on silhouette analysis to study the separation distance between the resulting clusters. We set the maximum number of clusters to $15$ and compute so-called silhouette coefficients for all of them. For all draws, we choose the optimal number of clusters based on this coefficient, which yields an empirical distribution of the number of clusters. Our findings are displayed in Table \ref{tab:clusterprobs}. The procedure selects $k=2$ in $86.9$ percent of the draws, and more clusters than $k=5$ are never supported. Consequently, we choose $k=2$ for all subsequent analyses. 

\begin{table*}[!htbp]
\caption{Identifying the number of clusters.}\vspace*{-1.5em}
\begin{center}
\begin{footnotesize}
\begin{threeparttable}
\begin{tabular*}{\textwidth}{@{\extracolsep{\fill}} lcccc}
  \toprule
\textbf{Number of clusters} & 2 & 3 & 4 & 5 \\
  \midrule
  Probability (\%) & 86.9 & 8.5 & 2.4 & 2.2 \\ 
   \bottomrule
\end{tabular*}
\begin{tablenotes}[para,flushleft]
\scriptsize{\textit{Notes}: Industries are clustered based on total and network effects per industry. We use silhouette analysis for all posterior draws using a maximum value of $15$ clusters. The number of clusters is selected based on the so-called silhouette coefficient, which yields an empirical distribution for the most adequate number of clusters.}
\end{tablenotes}
\end{threeparttable}
\end{footnotesize}
\end{center}
\label{tab:clusterprobs}
\end{table*}

\begin{figure}[t]
\includegraphics[width=\textwidth]{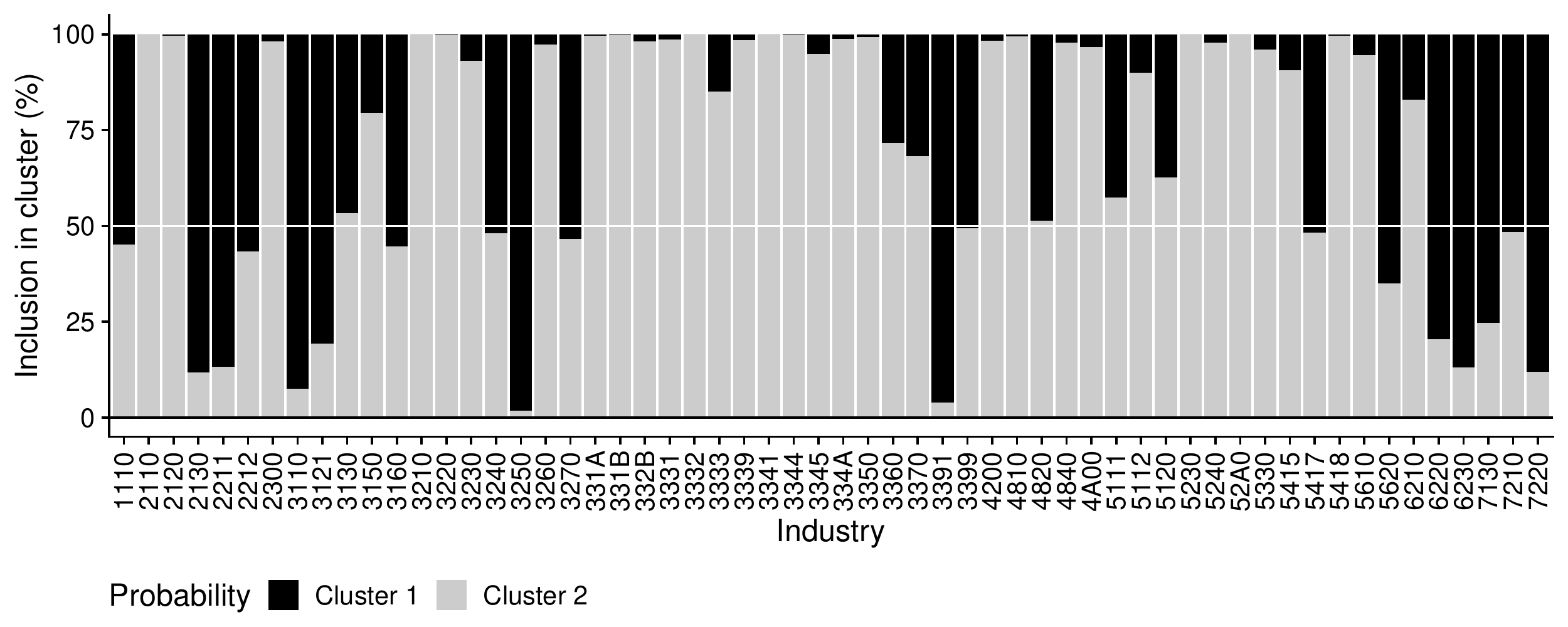}
\caption{Cluster allocation of industries.}\label{fig:industry-cluster}\vspace*{-0.3cm}
\caption*{\footnotesize\textit{Note}: The number of clusters is chosen to be $k=2$ based on silhouette analysis. Indicated values are empirical inclusion probabilities for industries in clusters across posterior draws. The white line marks the $50$ percent threshold.}
\end{figure}

\begin{figure}[t]
\includegraphics[width=\textwidth]{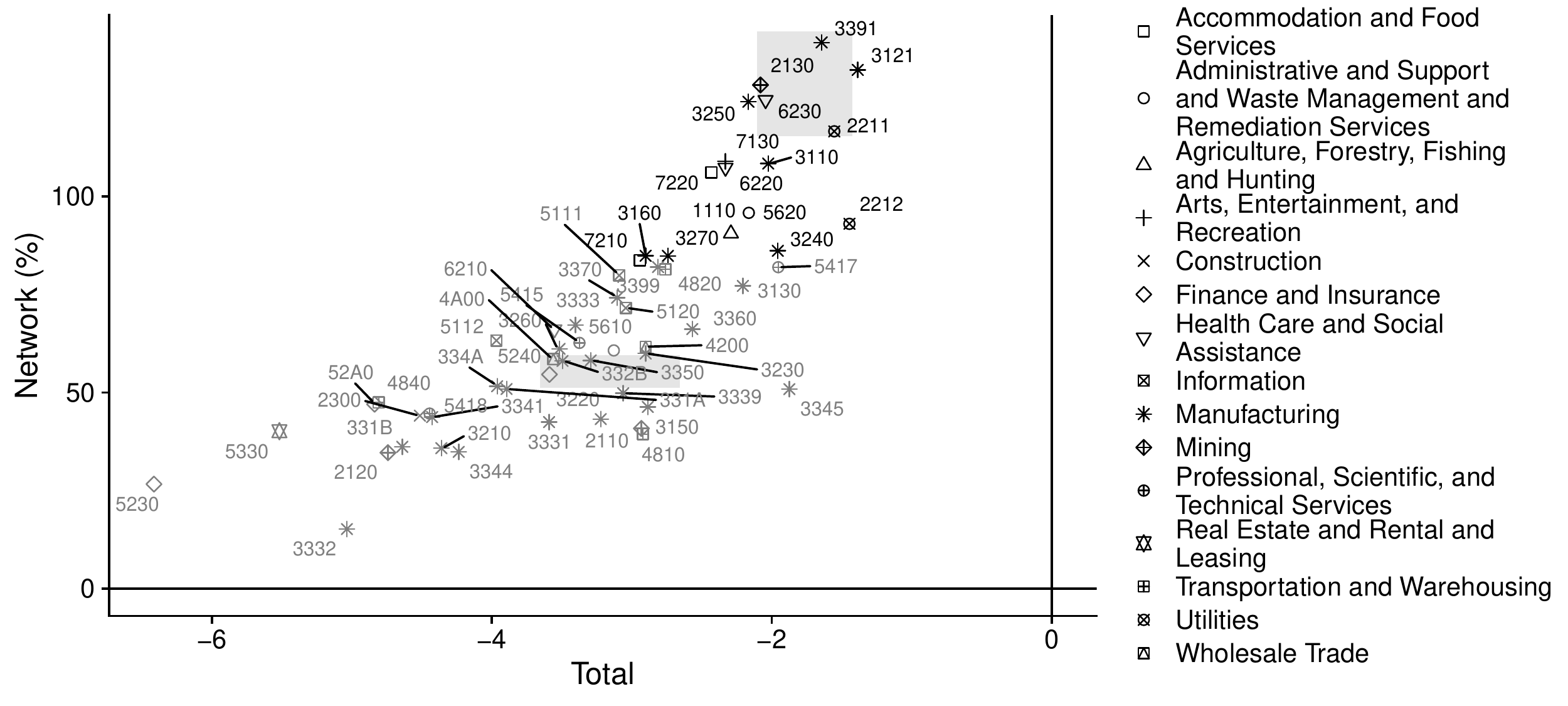}
\caption{Effect sizes and clustering of industries.}\label{fig:industry-scatter}\vspace*{-0.3cm}
\caption*{\footnotesize\textit{Note}: Points indicate the posterior median of the indicated effect across industries. Industry-categories are based on the two-digit level NAICS codes. The number of clusters is chosen to be $k=2$ based on silhouette analysis. The grey shaded areas mark the empirical distribution of the cluster centers across all posterior draws.}
\end{figure}

The procedure outlined above produces empirical inclusion probabilities in clusters for all industries, across posterior draws.\footnote{Note that clusters are subject to identification issues \citep[see][]{fruhwirth2006finite}. We solve these by imposing an ordering constraint such that for each draw, the mean of network effects in Cluster $1$ is always larger than in Cluster $2$.} The findings for this exercise are summarized in Figure \ref{fig:industry-cluster}. To provide a more detailed interpretation of the obtained clusters, Figure \ref{fig:industry-scatter} shows a scatter plot between the posterior median of network and total effects. Industry-categories are based on the two-digit level NAICS codes. The grey shaded areas mark the empirical distribution of the cluster centers across all posterior draws. 

The clusters are of different sizes, and Cluster $1$ features less observations than Cluster $2$. For industries assigned to Cluster $1$, probabilities are often close to $50$ percent, indicating that membership assignment is fuzzy. Assessing the means of the estimated clusters in Figure \ref{fig:industry-scatter}, we find that Cluster $1$ is characterized by high network (exceeding $100$ percent) and comparatively small total effects (just below $-2$), while Cluster $2$ exhibits larger total effects (albeit with larger variance across industries) and network effects of about $55$ percent. Interestingly, we find a negative correlation between total effect sizes in absolute value and the strength of network effects per industry.

Zooming in on industry-characteristics in the context of our clustering analysis, several findings are worth noting. First, there is no clear-cut assignment of industries by their aggregate category. We can explain this finding by the respective closeness to end-consumers of industries. Monetary policy shocks in our framework are interpreted as demand shocks, which implies that industries that are closer to end-consumers are affected directly, while these effects are transmitted upstream via network effects to the suppliers of these industries in the production network. An illustrative example is ``Securities, Commodity Contracts, and Other Financial Investments and Related Activities (5230),'' with a small magnitude of network effects, but large total impacts. Second, with some exceptions, most manufacturing industries are located in Cluster $1$, indicating comparatively low network effects. Assessing the manufacturing industries associated with Cluster $2$ in detail, we find that these are mainly industries located further up the supply chain (based on calculations using IO-tables), such as ``Food/Beverage manufacturing (3110/3121),'' or ``Medical Equipment and Supplies Manufacturing (3391).''

\begin{figure}[!htbp]
\includegraphics[width=\textwidth]{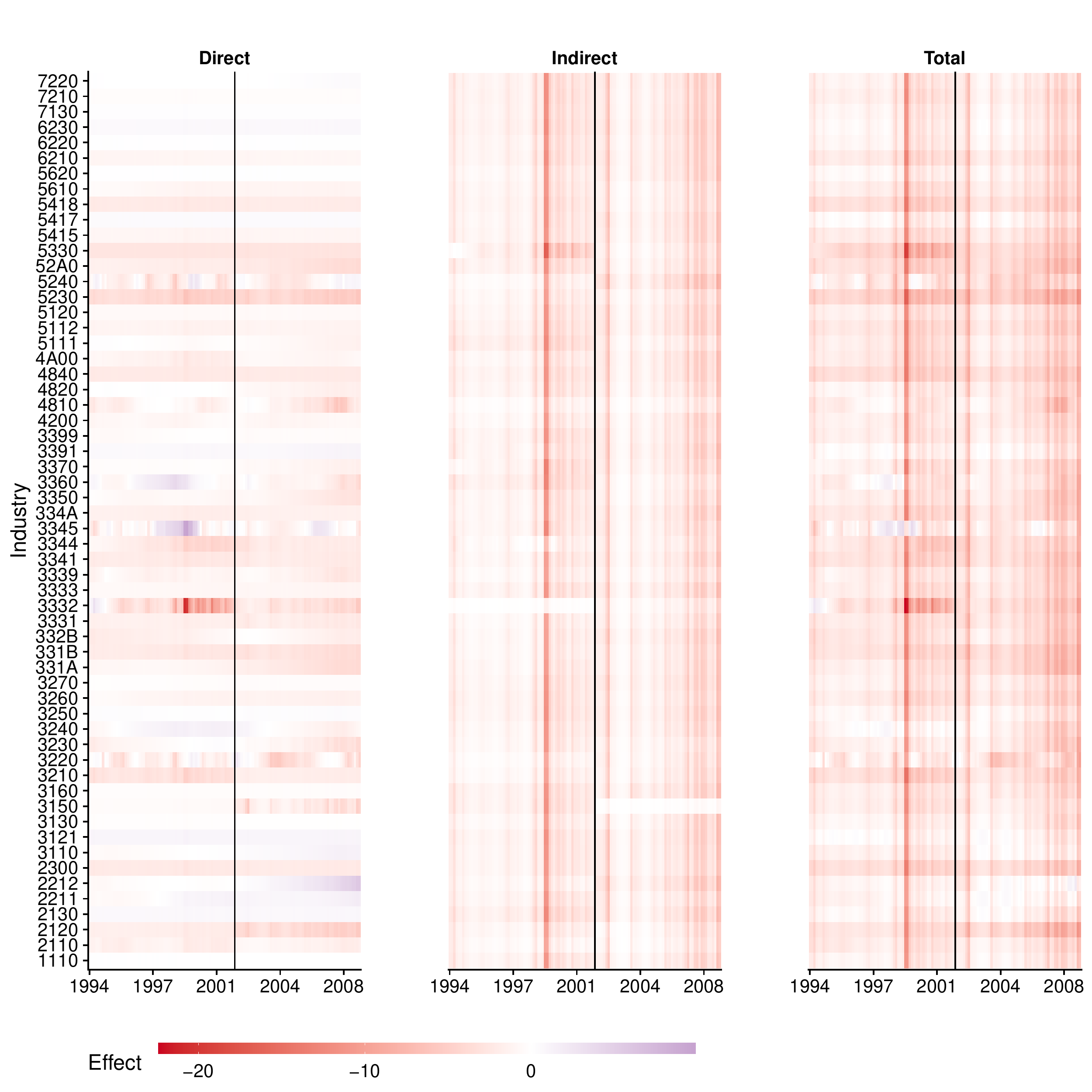}
\caption{Industry effects over time.}\label{fig:fx-heat}\vspace*{-0.3cm}
\caption*{\footnotesize\textit{Note}: Details on the impact measures are described in Section \ref{subsec:interpretation}. The heatmap shows the estimated posterior median impacts across industries and over time. The vertical black solid line indicates the January 30th 2002 policy meeting, where the weights matrix changes.}
\end{figure}

Turning to industry effects over time and the cross-section, Figure \ref{fig:fx-heat} shows posterior estimates of direct, indirect and total effects. Here, we again observe several noteworthy patterns. First, average patterns of differences of the impacts over time addressed previously are clearly visible in the industry-specific plots. The peak between the years 2002 and 2003 is clearly featured in all industries, while the gradual increase of monetary policy effects towards the end of the sample is visible. Second, a substantial share of industries shows small or even positive direct effects. Even though some direct effects are positive, total effects are for the most part, as expected, negative. This is mainly driven by the higher-order network effects. These findings relate directly to our previous discussion of industry clusters and closeness to end-consumers as determinants of the share of network effects, in line with \citet{ozdagli2017monetary}. Third, we detect several differences in industry-specific effects over time. Starting with direct effects, there are some industries such as ``Securities, Commodity Contracts, and Other Financial Investments and Related Activities (5230)'' or several of the manufacturing industries where we observe persistently strong or weak direct effects. By contrast, high-frequency movements are for instance observable in ``Industrial Machinery Manufacturing (3332),'' while in general, higher frequency movements in total effects are almost exclusively driven by indirect effects. Finally, there appears to be a break in the relative importance of industries in the production network governed by the network structure in $\bm{W}_t$. In January 2002, when the weights matrix is updated, we find that indirect effect patterns change for some industries. Examples are ``Industrial Machinery Manufacturing (3332),'' where indirect effects played only a minor role up to this date, or ``Apparel Manufacturing (3150),'' where after 2002 indirect effects are muted. It is worth mentioning that this break is not visible in the network dependence parameter or the effects averaged across industries.

\section{Closing remarks}\label{sec:conclusions}
This paper studies the impact of monetary policy on stock returns. We propose a novel Bayesian network panel state-space model to capture the propagation of shocks through the US production network. Alongside TVPs, our model addresses time-varying higher-order effects of monetary policy. Our results suggest substantial differences in industry responses that also vary significantly over time. We identify periods featuring increased economic and financial uncertainty, and periods when credit market conditions are tight as those where the impact of monetary policy actions is amplified. Moreover, our results suggest that policy responses in the US production network can be characterized by two main clusters. The clusters can be related to the closeness to end-consumers of the respective industries.

\small{\setstretch{0.85}
\addcontentsline{toc}{section}{References}
\bibliographystyle{custom.bst}
\bibliography{lit}}\normalsize

\clearpage
\begin{center}
\LARGE\textbf{Appendices}
\end{center}

\begin{appendices}\crefalias{section}{appsec}
\setcounter{equation}{0}
\renewcommand\theequation{A.\arabic{equation}}
\section{MCMC algorithm}\label{app:algorithm}
We use the following algorithm to generate draws for all parameters of the model by a standard MCMC sampling algorithm. Specifically, the sampler iterates through the following steps:
\begin{enumerate}[align=left]
	\item Conditional on all other parameters of the model, the time-varying regression coefficients are simulated independently on an industry-by-industry basis using an FFBS algorithm \citep{doi:10.1093/biomet/81.3.541,doi:10.1111/j.1467-9892.1994.tb00184.x}.
	\item Given the full history of the time-varying parameters $\{\bm{\tilde\theta}_{it}\}_{t=1}^T$, the initial state $\bm{\theta}_{i0}$ and the square root of the state innovation variances $\omega_{i1},\hdots,\sqrt{\omega_{iK+1}}$ are drawn in one block from their Gaussian posterior distribution, see \citet{FRUHWIRTHSCHNATTER201085}.
	\item The measurement equation error variances $\sigma_i^{2}$ are drawn from their inverse Gamma conditional posterior distributions again on an industry-by-industry basis. The posterior moments can be found, for instance, in \citet{koop2003bayesian}.
	\item The full history of the network dependence parameter $\{\rho_t\}_{t=1}^T$ conditional on all other model parameters is simulated using the Metropolis-Hastings algorithm discussed in Section \ref{sec:posterior}. The algorithm involves proposing new values for $\rho_t$ at each point in time. These values are subsequently evaluated and used for constructing acceptance probabilities. 
	\item Conditional on $\{\rho_t\}_{t=1}^T$, the state innovation variances for the network dependence parameter are simulated from their inverse Gamma posterior distribution, with the moments corresponding to a standard linear regression model \citep[see][]{koop2003bayesian}.
\end{enumerate}

This completes the MCMC algorithm employed to simulate from the posterior distribution. After choosing starting values and a sufficient burn-in period we store draws from the conditional posterior distributions. In particular, we discard the initial $5,000$ draws, while Bayesian inference is performed based on every second of the subsequent $10,000$ draws resulting in a set of $5,000$ draws from the posterior. For the sake of brevity, we only report posterior estimates of parameters and higher-order functions of them that are of direct interest. Additional results are available upon request.

The sampler takes about $37$ minutes to produce the $15,000$ draws in the case of the most flexible specification on a 2016 Macbook Pro with 2.9GHz Dual-Core Intel Core i5 with 8GB RAM running \texttt{R 4.0.0}. This runtime excludes the construction of the impact matrix $\bm{S}_{kt}$ which can be quite time-consuming due to the dimensionality of the underlying panel data. However, this step can be performed outside of the main sampling loop.

\clearpage
\renewcommand\theequation{B.\arabic{equation}}
\section{Data}\label{app:data}
All data and replication files are available from the authors upon request. Table \ref{tab:industries} shows the four-digit NAICS codes alongside of a description of the industry and categories derived from two-digit level codes for the aggregation scheme in the context of the 1997 and 2002 IO-tables.

\begin{table}[!htbp]
\centering\tiny
\caption{List of industries.}\vspace*{-0.5em}
\begin{threeparttable}
\begin{tabular}{lp{9.5cm}p{6cm}}
  \toprule
\textbf{NAICS} & \textbf{Description} & \textbf{Category}\\ 
  \midrule
1110 & Crop Production            & Agriculture, Forestry, Fishing and Hunting         \\ 
  2110 & Oil and Gas Extraction          & Mining             \\ 
  2120 & Mining (except Oil and Gas)         & Mining             \\ 
  2130 & Support Activities for Mining          & Mining             \\ 
  2211 & Electric Power Generation, Transmission and Distribution        & Utilities             \\ 
  2212 & Natural Gas Distribution           & Utilities             \\ 
  2300 & Construction (Miscellaneous) & Construction \\ 
  3110 & Food Manufacturing            & Manufacturing \\ 
  3121 & Beverage Manufacturing            & Manufacturing \\ 
  3130 & Textile Mills            & Manufacturing \\ 
  3150 & Apparel Manufacturing            & Manufacturing \\ 
  3160 & Leather and Allied Product Manufacturing         & Manufacturing \\ 
  3210 & Wood Product Manufacturing           & Manufacturing \\ 
  3220 & Paper Manufacturing            & Manufacturing \\ 
  3230 & Printing and Related Support Activities         & Manufacturing \\ 
  3240 & Petroleum and Coal Products Manufacturing         & Manufacturing \\ 
  3250 & Chemical Manufacturing            & Manufacturing \\ 
  3260 & Plastics and Rubber Products Manufacturing         & Manufacturing \\ 
  3270 & Nonmetallic Mineral Product Manufacturing          & Manufacturing \\ 
  3331 & Agriculture, Construction, and Mining Machinery Manufacturing        & Manufacturing \\ 
  3332 & Industrial Machinery Manufacturing           & Manufacturing \\ 
  3333 & Commercial and Service Industry Machinery Manufacturing        & Manufacturing \\ 
  3339 & Other General Purpose Machinery Manufacturing         & Manufacturing \\ 
  3341 & Computer and Peripheral Equipment Manufacturing         & Manufacturing \\ 
  3344 & Semiconductor and Other Electronic Component Manufacturing        & Manufacturing \\ 
  3345 & Navigational, Measuring, Electromedical, and Control Instruments Manufacturing       & Manufacturing \\ 
  3350 & Electrical Equipment, Appliance, and Component Manufacturing        & Manufacturing \\ 
  3360 & Transportation Equipment Manufacturing           & Manufacturing \\ 
  3370 & Furniture and Related Product Manufacturing         & Manufacturing \\ 
  3391 & Medical Equipment and Supplies Manufacturing         & Manufacturing \\ 
  3399 & Other Miscellaneous Manufacturing           & Manufacturing \\ 
  331A & Primary Metal Manufacturing (A) & Manufacturing \\ 
  331B & Primary Metal Manufacturing (B) & Manufacturing \\ 
  332B & Fabricated Metal Product Manufacturing (B) & Manufacturing \\ 
  334A & Computer and Electronic Product Manufacturing (A) & Manufacturing \\ 
  4200 & Wholesale Trade (Miscellaneous) & Wholesale Trade \\ 
  4A00 & Commercial (Miscellaneous) & Wholesale Trade \\ 
  4810 & Air Transportation            & Transportation and Warehousing \\ 
  4820 & Rail Transportation            & Transportation and Warehousing \\ 
  4840 & Truck Transportation            & Transportation and Warehousing \\ 
  5111 & Newspaper, Periodical, Book, and Directory Publishers        & Information             \\ 
  5112 & Software Publishers            & Information             \\ 
  5120 & Motion Picture and Sound Recording Industries        & Information             \\ 
  5230 & Securities, Commodity Contracts, and Other Financial Investments and Related Activities    & Finance and Insurance           \\ 
  5240 & Insurance Carriers and Related Activities         & Finance and Insurance           \\ 
  52A0 & Finance and Insurance (Miscellaneous) & Finance and Insurance   \\ 
  5330 & Lessors of Nonfinancial Intangible Assets (except Copyrighted Works)      & Real Estate and Rental and Leasing        \\ 
  5415 & Computer Systems Design and Related Services        & Professional, Scientific, and Technical Services         \\ 
  5417 & Scientific Research and Development Services         & Professional, Scientific, and Technical Services         \\ 
  5418 & Advertising and Related Services          & Professional, Scientific, and Technical Services         \\ 
  5610 & Administrative and Support Services          & Administrative and Support and Waste Management and Remediation Services     \\ 
  5620 & Waste Management and Remediation Services         & Administrative and Support and Waste Management and Remediation Services     \\ 
  6210 & Ambulatory Health Care Services          & Health Care and Social Assistance         \\ 
  6220 & Hospitals             & Health Care and Social Assistance         \\ 
  6230 & Nursing and Residential Care Facilities         & Health Care and Social Assistance         \\ 
  7130 & Amusement, Gambling, and Recreation Industries         & Arts, Entertainment, and Recreation          \\ 
  7210 & Accommodation             & Accommodation and Food Services          \\ 
  7220 & Food Services and Drinking Places         & Accommodation and Food Services    \\
   \bottomrule
\end{tabular}
\begin{tablenotes}[para,flushleft]
\tiny{\textit{Notes}: ``NAICS'' gives the industry classification code, ``Description'' is the name of the respective industry. ``Category'' provides summary aggregates of industries using the two-digit level codes.}
\end{tablenotes}
\end{threeparttable}
\label{tab:industries}
\end{table}

\clearpage
\renewcommand\theequation{C.\arabic{equation}}
\section{Robustness analysis}\label{app:robustness}
The following tables provide additional results for weights matrices and industry aggregations based on 1992, 1997 and 2002 IO-tables. Moreover, we consider a split-sample analysis, where the full sampling period is partitioned into $60$ policy meetings each. 

\subsection{Different weights matrices and aggregation schemes}
\begin{table*}[htbp]
\caption{Estimated impacts of monetary policy on stock returns across industries, 1992 IO-tables.}\vspace*{-1.5em}
\begin{center}
\begin{tiny}
\begin{threeparttable}
\begin{tabular*}{\textwidth}{@{\extracolsep{\fill}} lcccccccc}
\toprule
               & \multicolumn{4}{c}{\textbf{Parameters}} & \multicolumn{4}{c}{\textbf{Impacts}}\\
               \cmidrule(lr){2-5}\cmidrule(lr){6-9}
   & $\alpha$ & $\beta$ & $\sigma^2$ & $\rho$ & Indirect & Direct & Total & Netw. (\%) \\ 
\midrule
  B1 & -0.11 & -2.99 & 0.15 &  &  & -2.99 & -2.99 &  \\ 
   & (-0.13, -0.09) & (-3.25, -2.72) & (0.14, 0.15) &  &  & (-3.25, -2.72) & (-3.25, -2.72) &  \\ 
  B2 & -0.11 & -2.83 & 0.31 &  &  & -2.83 & -2.83 &  \\ 
   & (-0.13, -0.09) & (-3.08, -2.59) & (0.30, 0.33) &  &  & (-3.08, -2.59) & (-3.08, -2.59) &  \\ 
  B3 & -0.03 & -0.79 & 0.10 & 0.74 & -1.95 & -1.15 & -3.10 & 62.9 \\ 
   & (-0.04, -0.02) & (-0.96, -0.64) & (0.10, 0.11) & (0.72, 0.76) & (-2.33, -1.59) & (-1.38, -0.92) & (-3.69, -2.51) & (60.7, 65.0) \\ 
  B4 & -0.01 & -0.35 & 0.09 & 0.89 & -2.33 & -0.57 & -2.90 & 80.4 \\ 
   & (-0.02,  0.00) & (-0.48, -0.20) & (0.08, 0.10) & (0.87, 0.90) & (-3.16, -1.38) & (-0.76, -0.34) & (-3.87, -1.77) & (77.6, 83.0) \\ 
   \midrule
  C1 & -0.04 & -0.87 & 0.11 & 0.60 & -1.67 & -1.17 & -2.84 & 58.8 \\ 
   & (-0.05, -0.03) & (-1.08, -0.66) & (0.10, 0.11) & (0.57, 0.62) & (-2.33, -1.28) & (-1.44, -0.89) & (-3.60, -2.21) & (55.6, 67.1) \\ 
  C2 & -0.02 & -0.45 & 0.09 & 0.79 & -2.25 & -0.66 & -2.91 & 77.2 \\ 
   & (-0.03, -0.01) & (-0.60, -0.28) & (0.09, 0.10) & (0.77, 0.81) & (-3.55, -1.49) & (-0.90, -0.44) & (-4.37, -1.96) & (73.0, 83.4) \\ 
  C3 & -0.11 & -2.86 & 0.31 &  &  & -2.86 & -2.86 &  \\ 
   & (-0.13, -0.09) & (-3.10, -2.61) & (0.29, 0.33) &  &  & (-3.10, -2.61) & (-3.10, -2.61) &  \\ 
  C4 & -0.01 & -0.36 & 0.09 & 0.90 & -3.19 & -0.69 & -3.88 & 82.3 \\ 
   & (-0.02,  0.00) & (-0.50, -0.22) & (0.08, 0.10) & (0.89, 0.91) & (-4.18, -2.25) & (-0.89, -0.49) & (-5.12, -2.74) & (80.2, 84.4) \\ 
  C5 & -0.01 & -0.62 & 0.09 & 0.82 & -4.35 & -1.05 & -5.41 & 80.4 \\ 
   & (-0.02,  0.00) & (-0.79, -0.47) & (0.08, 0.10) & (0.81, 0.84) & (-7.92, -2.87) & (-1.34, -0.81) & (-9.01, -3.71) & (76.5, 87.7) \\ 
\bottomrule
\end{tabular*}
\begin{tablenotes}[para,flushleft]
\tiny{\textit{Notes}: For model specifications see Table \ref{tab:models}. For those featuring heterogeneous coefficients over the cross-section $i$ or over time $t$, we take the arithmetic mean over all industries and over time per iteration of the algorithm and report the resulting posterior percentiles (the posterior median, and the bounds in parentheses marking the $99$ percent posterior credible set).}
\end{tablenotes}
\end{threeparttable}
\end{tiny}
\end{center}
\label{tab:results_rob1992}
\end{table*}

\begin{table*}[htbp]
\caption{Estimated impacts of monetary policy on stock returns across industries, 1997 IO-tables.}\vspace*{-1.5em}
\begin{center}
\begin{tiny}
\begin{threeparttable}
\begin{tabular*}{\textwidth}{@{\extracolsep{\fill}} lcccccccc}
\toprule
               & \multicolumn{4}{c}{\textbf{Parameters}} & \multicolumn{4}{c}{\textbf{Impacts}}\\
               \cmidrule(lr){2-5}\cmidrule(lr){6-9}
   & $\alpha$ & $\beta$ & $\sigma^2$ & $\rho$ & Indirect & Direct & Total & Netw. (\%) \\ 
\midrule
  B1 & -0.13 & -3.14 & 0.15 &  &  & -3.14 & -3.14 &  \\ 
   & (-0.14, -0.11) & (-3.42, -2.88) & (0.14, 0.15) &  &  & (-3.42, -2.88) & (-3.42, -2.88) &  \\ 
  B2 & -0.12 & -2.97 & 0.34 &  &  & -2.97 & -2.97 &  \\ 
   & (-0.14, -0.10) & (-3.21, -2.73) & (0.32, 0.35) &  &  & (-3.21, -2.73) & (-3.21, -2.73) &  \\ 
  B3 & -0.04 & -1.07 & 0.15 & 0.66 & -1.71 & -1.41 & -3.12 & 54.7 \\ 
   & (-0.06, -0.03) & (-1.28, -0.89) & (0.14, 0.15) & (0.63, 0.68) & (-2.02, -1.41) & (-1.68, -1.17) & (-3.66, -2.57) & (52.1, 57.2) \\ 
  B4 & -0.01 & -0.30 & 0.12 & 0.90 & -2.49 & -0.58 & -3.08 & 81.2 \\ 
   & (-0.02,  0.00) & (-0.48, -0.13) & (0.11, 0.13) & (0.88, 0.91) & (-3.52, -1.37) & (-0.80, -0.33) & (-4.20, -1.79) & (77.9, 84.2) \\ 
   \midrule
  C1 & -0.05 & -1.02 & 0.15 & 0.57 & -1.34 & -1.29 & -2.63 & 51.1 \\ 
   & (-0.06, -0.04) & (-1.24, -0.79) & (0.14, 0.16) & (0.54, 0.59) & (-1.64, -1.06) & (-1.55, -1.01) & (-3.19, -2.11) & (48.3, 54.1) \\ 
  C2 & -0.02 & -0.35 & 0.12 & 0.83 & -2.22 & -0.60 & -2.84 & 78.6 \\ 
   & (-0.03, -0.01) & (-0.53, -0.16) & (0.12, 0.14) & (0.81, 0.85) & (-3.44, -1.27) & (-0.84, -0.34) & (-4.18, -1.66) & (74.6, 83.9) \\ 
  C3 & -0.12 & -2.97 & 0.34 &  &  & -2.97 & -2.97 &  \\ 
   & (-0.14, -0.11) & (-3.22, -2.70) & (0.32, 0.35) &  &  & (-3.22, -2.70) & (-3.22, -2.70) &  \\ 
  C4 & -0.01 & -0.28 & 0.12 & 0.90 & -2.63 & -0.58 & -3.21 & 82.1 \\ 
   & (-0.02,  0.00) & (-0.45, -0.11) & (0.11, 0.13) & (0.88, 0.92) & (-3.79, -1.58) & (-0.79, -0.36) & (-4.57, -1.96) & (78.9, 84.9) \\ 
  C5 & -0.02 & -0.34 & 0.12 & 0.84 & -2.44 & -0.61 & -3.05 & 79.8 \\ 
   & (-0.03, -0.01) & (-0.53, -0.15) & (0.11, 0.14) & (0.82, 0.86) & (-3.82, -1.38) & (-0.88, -0.37) & (-4.54, -1.78) & (75.6, 84.9) \\ 
\bottomrule
\end{tabular*}
\begin{tablenotes}[para,flushleft]
\tiny{\textit{Notes}: For model specifications see Table \ref{tab:models}. For those featuring heterogeneous coefficients over the cross-section $i$ or over time $t$, we take the arithmetic mean over all industries and over time per iteration of the algorithm and report the resulting posterior percentiles (the posterior median, and the bounds in parentheses marking the $99$ percent posterior credible set).}
\end{tablenotes}
\end{threeparttable}
\end{tiny}
\end{center}
\label{tab:results_rob1997}
\end{table*}

\begin{table*}[htbp]
\caption{Estimated impacts of monetary policy on stock returns across industries, 2002 IO-tables.}\vspace*{-1.5em}
\begin{center}
\begin{tiny}
\begin{threeparttable}
\begin{tabular*}{\textwidth}{@{\extracolsep{\fill}} lcccccccc}
\toprule
               & \multicolumn{4}{c}{\textbf{Parameters}} & \multicolumn{4}{c}{\textbf{Impacts}}\\
               \cmidrule(lr){2-5}\cmidrule(lr){6-9}
   & $\alpha$ & $\beta$ & $\sigma^2$ & $\rho$ & Indirect & Direct & Total & Netw. (\%) \\ 
\midrule
  B1 & -0.12 & -3.14 & 0.15 &  &  & -3.14 & -3.14 &  \\ 
   & (-0.13, -0.10) & (-3.41, -2.87) & (0.14, 0.15) &  &  & (-3.41, -2.87) & (-3.41, -2.87) &  \\ 
  B2 & -0.11 & -2.98 & 0.33 &  &  & -2.98 & -2.98 &  \\ 
   & (-0.13, -0.10) & (-3.23, -2.68) & (0.31, 0.35) &  &  & (-3.23, -2.68) & (-3.23, -2.68) &  \\ 
  B3 & -0.04 & -0.96 & 0.14 & 0.69 & -1.84 & -1.27 & -3.12 & 59.1 \\ 
   & (-0.05, -0.02) & (-1.16, -0.77) & (0.13, 0.14) & (0.67, 0.71) & (-2.23, -1.50) & (-1.53, -1.03) & (-3.72, -2.52) & (56.7, 61.3) \\ 
  B4 & -0.01 & -0.40 & 0.12 & 0.87 & -2.44 & -0.66 & -3.10 & 78.5 \\ 
   & (-0.03,  0.00) & (-0.58, -0.24) & (0.11, 0.13) & (0.85, 0.88) & (-3.34, -1.47) & (-0.88, -0.45) & (-4.20, -1.89) & (75.8, 80.9) \\ 
   \midrule
  C1 & -0.05 & -1.09 & 0.14 & 0.52 & -1.58 & -1.35 & -2.93 & 54.1 \\ 
   & (-0.06, -0.04) & (-1.32, -0.86) & (0.13, 0.15) & (0.49, 0.55) & (-2.48, -1.23) & (-1.62, -1.07) & (-3.90, -2.32) & (50.6, 63.3) \\ 
  C2 & -0.02 & -0.48 & 0.12 & 0.77 & -2.18 & -0.72 & -2.90 & 75.1 \\ 
   & (-0.03, -0.01) & (-0.66, -0.29) & (0.11, 0.14) & (0.74, 0.79) & (-3.44, -1.33) & (-0.94, -0.48) & (-4.31, -1.83) & (71.7, 80.7) \\ 
  C3 & -0.12 & -3.00 & 0.33 &  &  & -3.00 & -3.00 &  \\ 
   & (-0.13, -0.10) & (-3.26, -2.74) & (0.31, 0.35) &  &  & (-3.26, -2.74) & (-3.26, -2.74) &  \\ 
  C4 & -0.01 & -0.41 & 0.12 & 0.87 & -2.68 & -0.71 & -3.40 & 79.0 \\ 
   & (-0.03,  0.00) & (-0.58, -0.23) & (0.11, 0.13) & (0.86, 0.88) & (-3.56, -1.78) & (-0.93, -0.51) & (-4.47, -2.33) & (76.5, 81.2) \\ 
  C5 & -0.02 & -0.58 & 0.12 & 0.80 & -3.06 & -0.92 & -3.98 & 76.9 \\ 
   & (-0.03, -0.01) & (-0.79, -0.40) & (0.11, 0.13) & (0.78, 0.82) & (-4.78, -1.97) & (-1.20, -0.67) & (-5.81, -2.65) & (72.0, 82.5) \\ 
\bottomrule
\end{tabular*}
\begin{tablenotes}[para,flushleft]
\tiny{\textit{Notes}: For model specifications see Table \ref{tab:models}. For those featuring heterogeneous coefficients over the cross-section $i$ or over time $t$, we take the arithmetic mean over all industries and over time per iteration of the algorithm and report the resulting posterior percentiles (the posterior median, and the bounds in parentheses marking the $99$ percent posterior credible set).}
\end{tablenotes}
\end{threeparttable}
\end{tiny}
\end{center}
\label{tab:results_rob2002}
\end{table*}

\clearpage
\subsection{Split-sample analysis, 1994--2001 (first 60 FOMC meetings)}
\begin{table*}[htbp]
\caption{Estimated impacts of monetary policy on stock returns across industries, 1992 IO-tables.}\vspace*{-1.5em}
\begin{center}
\begin{tiny}
\begin{threeparttable}
\begin{tabular*}{\textwidth}{@{\extracolsep{\fill}} lcccccccc}
\toprule
               & \multicolumn{4}{c}{\textbf{Parameters}} & \multicolumn{4}{c}{\textbf{Impacts}}\\
               \cmidrule(lr){2-5}\cmidrule(lr){6-9}
   & $\alpha$ & $\beta$ & $\sigma^2$ & $\rho$ & Indirect & Direct & Total & Netw. (\%) \\ 
\midrule
A1 & -0.11 & -2.27 & 0.19 &  &  & -2.27 & -2.27 &  \\ 
   & (-0.26, 0.03) & (-4.53, 0.09) & (0.13, 0.32) &  &  & (-4.53, 0.09) & (-4.53, 0.09) &  \\ 
  A2 & -0.12 & -3.10 & 0.18 &  &  & -3.10 & -3.10 &  \\ 
   & (-0.27,  0.02) & (-5.65, -0.61) & (0.11, 0.31) &  &  & (-5.65, -0.61) & (-5.65, -0.61) &  \\ 
   \midrule
  B1 & -0.08 & -2.08 & 0.23 &  &  & -2.08 & -2.08 &  \\ 
   & (-0.11, -0.06) & (-2.45, -1.74) & (0.22, 0.25) &  &  & (-2.45, -1.74) & (-2.45, -1.74) &  \\ 
  B2 & -0.08 & -1.97 & 0.33 &  &  & -1.97 & -1.97 &  \\ 
   & (-0.11, -0.06) & (-2.31, -1.65) & (0.30, 0.38) &  &  & (-2.31, -1.65) & (-2.31, -1.65) &  \\ 
  B3 & -0.04 & -0.94 & 0.17 & 0.57 & -0.98 & -1.20 & -2.19 & 45.1 \\ 
   & (-0.05, -0.02) & (-1.24, -0.64) & (0.16, 0.19) & (0.53, 0.61) & (-1.31, -0.66) & (-1.57, -0.82) & (-2.83, -1.53) & (41.1, 49.3) \\ 
  B4 & -0.01 & -0.38 & 0.14 & 0.83 & -1.44 & -0.54 & -1.98 & 72.5 \\ 
   & (-0.03,  0.00) & (-0.63, -0.13) & (0.12, 0.17) & (0.80, 0.85) & (-2.30, -0.51) & (-0.86, -0.19) & (-3.11, -0.69) & (66.9, 78.2) \\ 
   \midrule
  C1 & -0.06 & -1.26 & 0.18 & 0.43 & -1.06 & -1.52 & -2.58 & 41.2 \\ 
   & (-0.08, -0.04) & (-1.59, -0.94) & (0.17, 0.19) & (0.37, 0.48) & (-1.45, -0.77) & (-1.90, -1.14) & (-3.26, -1.90) & (36.3, 48.5) \\ 
  C2 & -0.02 & -0.57 & 0.15 & 0.73 & -1.76 & -0.79 & -2.55 & 69.0 \\ 
   & (-0.04, -0.01) & (-0.83, -0.35) & (0.13, 0.18) & (0.70, 0.77) & (-3.14, -1.04) & (-1.12, -0.48) & (-4.18, -1.51) & (62.6, 78.6) \\ 
  C3 & -0.08 & -2.00 & 0.33 &  &  & -2.00 & -2.00 &  \\ 
   & (-0.11, -0.06) & (-2.35, -1.63) & (0.30, 0.37) &  &  & (-2.35, -1.63) & (-2.35, -1.63) &  \\ 
  C4 & -0.01 & -0.38 & 0.14 & 0.85 & -1.68 & -0.56 & -2.25 & 74.7 \\ 
   & (-0.03,  0.00) & (-0.60, -0.13) & (0.12, 0.17) & (0.82, 0.87) & (-2.62, -0.58) & (-0.85, -0.23) & (-3.45, -0.80) & (68.3, 80.2) \\ 
  C5 & -0.02 & -0.62 & 0.15 & 0.76 & -2.12 & -0.87 & -2.99 & 70.7 \\ 
   & (-0.04,  0.00) & (-0.88, -0.36) & (0.13, 0.18) & (0.72, 0.80) & (-3.53, -1.15) & (-1.20, -0.54) & (-4.55, -1.76) & (65.0, 79.1) \\ 
\bottomrule
\end{tabular*}
\begin{tablenotes}[para,flushleft]
\tiny{\textit{Notes}: For model specifications see Table \ref{tab:models}. For those featuring heterogeneous coefficients over the cross-section $i$ or over time $t$, we take the arithmetic mean over all industries and over time per iteration of the algorithm and report the resulting posterior percentiles (the posterior median, and the bounds in parentheses marking the $99$ percent posterior credible set).}
\end{tablenotes}
\end{threeparttable}
\end{tiny}
\end{center}
\label{tab:results_split1992}
\end{table*}

\begin{table*}[htbp]
\caption{Estimated impacts of monetary policy on stock returns across industries, 1997 IO-tables.}\vspace*{-1.5em}
\begin{center}
\begin{tiny}
\begin{threeparttable}
\begin{tabular*}{\textwidth}{@{\extracolsep{\fill}} lcccccccc}
\toprule
               & \multicolumn{4}{c}{\textbf{Parameters}} & \multicolumn{4}{c}{\textbf{Impacts}}\\
               \cmidrule(lr){2-5}\cmidrule(lr){6-9}
   & $\alpha$ & $\beta$ & $\sigma^2$ & $\rho$ & Indirect & Direct & Total & Netw. (\%) \\ 
\midrule
  B1 & -0.11 & -2.29 & 0.23 &  &  & -2.29 & -2.29 &  \\ 
   & (-0.14, -0.09) & (-2.71, -1.93) & (0.22, 0.25) &  &  & (-2.71, -1.93) & (-2.71, -1.93) &  \\ 
  B2 & -0.11 & -2.16 & 0.36 &  &  & -2.16 & -2.16 &  \\ 
   & (-0.14, -0.09) & (-2.50, -1.75) & (0.33, 0.40) &  &  & (-2.50, -1.75) & (-2.50, -1.75) &  \\ 
  B3 & -0.06 & -1.21 & 0.23 & 0.47 & -0.84 & -1.43 & -2.29 & 37.1 \\ 
   & (-0.08, -0.04) & (-1.55, -0.91) & (0.22, 0.25) & (0.42, 0.51) & (-1.11, -0.65) & (-1.86, -1.09) & (-2.94, -1.75) & (33.2, 40.8) \\ 
  B4 & -0.03 & -0.41 & 0.19 & 0.81 & -1.55 & -0.65 & -2.21 & 70.1 \\ 
   & (-0.05, -0.01) & (-0.70, -0.13) & (0.17, 0.22) & (0.77, 0.84) & (-2.43, -0.65) & (-1.02, -0.32) & (-3.42, -1.00) & (64.5, 75.5) \\ 
   \midrule
  C1 & -0.07 & -1.43 & 0.23 & 0.39 & -0.84 & -1.65 & -2.48 & 33.9 \\ 
   & (-0.09, -0.05) & (-1.77, -1.10) & (0.22, 0.25) & (0.34, 0.43) & (-1.08, -0.63) & (-2.03, -1.29) & (-3.03, -1.95) & (29.6, 38.3) \\ 
  C2 & -0.03 & -0.60 & 0.19 & 0.73 & -1.79 & -0.89 & -2.68 & 66.7 \\ 
   & (-0.05, -0.01) & (-0.88, -0.32) & (0.17, 0.22) & (0.70, 0.77) & (-2.61, -1.04) & (-1.23, -0.54) & (-3.84, -1.59) & (61.2, 72.9) \\ 
  C3 & -0.11 & -2.16 & 0.36 &  &  & -2.16 & -2.16 &  \\ 
   & (-0.14, -0.09) & (-2.54, -1.80) & (0.34, 0.40) &  &  & (-2.54, -1.80) & (-2.54, -1.80) &  \\ 
  C4 & -0.02 & -0.40 & 0.19 & 0.81 & -1.63 & -0.67 & -2.31 & 70.9 \\ 
   & (-0.05, -0.01) & (-0.68, -0.11) & (0.17, 0.22) & (0.78, 0.85) & (-2.56, -0.78) & (-1.03, -0.32) & (-3.50, -1.13) & (65.1, 76.8) \\ 
  C5 & -0.03 & -0.58 & 0.19 & 0.74 & -1.89 & -0.89 & -2.79 & 67.9 \\ 
   & (-0.05, -0.01) & (-0.88, -0.27) & (0.17, 0.22) & (0.69, 0.78) & (-2.91, -0.96) & (-1.25, -0.51) & (-4.04, -1.46) & (61.2, 74.9) \\ 
\bottomrule
\end{tabular*}
\begin{tablenotes}[para,flushleft]
\tiny{\textit{Notes}: For model specifications see Table \ref{tab:models}. For those featuring heterogeneous coefficients over the cross-section $i$ or over time $t$, we take the arithmetic mean over all industries and over time per iteration of the algorithm and report the resulting posterior percentiles (the posterior median, and the bounds in parentheses marking the $99$ percent posterior credible set).}
\end{tablenotes}
\end{threeparttable}
\end{tiny}
\end{center}
\label{tab:results_split1997}
\end{table*}

\begin{table*}[htbp]
\caption{Estimated impacts of monetary policy on stock returns across industries, 2002 IO-tables.}\vspace*{-1.5em}
\begin{center}
\begin{tiny}
\begin{threeparttable}
\begin{tabular*}{\textwidth}{@{\extracolsep{\fill}} lcccccccc}
\toprule
               & \multicolumn{4}{c}{\textbf{Parameters}} & \multicolumn{4}{c}{\textbf{Impacts}}\\
               \cmidrule(lr){2-5}\cmidrule(lr){6-9}
   & $\alpha$ & $\beta$ & $\sigma^2$ & $\rho$ & Indirect & Direct & Total & Netw. (\%) \\ 
\midrule
  B1 & -0.10 & -2.29 & 0.23 &  &  & -2.29 & -2.29 &  \\ 
   & (-0.12, -0.08) & (-2.68, -1.89) & (0.22, 0.25) &  &  & (-2.68, -1.89) & (-2.68, -1.89) &  \\ 
  B2 & -0.10 & -2.18 & 0.36 &  &  & -2.18 & -2.18 &  \\ 
   & (-0.12, -0.07) & (-2.56, -1.79) & (0.32, 0.40) &  &  & (-2.56, -1.79) & (-2.56, -1.79) &  \\ 
  B3 & -0.05 & -1.10 & 0.21 & 0.52 & -0.99 & -1.32 & -2.31 & 42.7 \\ 
   & (-0.07, -0.03) & (-1.42, -0.80) & (0.20, 0.23) & (0.48, 0.57) & (-1.31, -0.74) & (-1.70, -0.97) & (-2.96, -1.69) & (38.7, 46.9) \\ 
  B4 & -0.02 & -0.49 & 0.19 & 0.79 & -1.56 & -0.70 & -2.27 & 69.0 \\ 
   & (-0.04,  0.00) & (-0.75, -0.25) & (0.16, 0.23) & (0.76, 0.82) & (-2.35, -0.81) & (-1.00, -0.39) & (-3.30, -1.19) & (64.3, 73.6) \\ 
   \midrule
  C1 & -0.07 & -1.51 & 0.22 & 0.36 & -1.08 & -1.72 & -2.80 & 38.6 \\ 
   & (-0.09, -0.05) & (-1.91, -1.11) & (0.20, 0.23) & (0.32, 0.41) & (-1.60, -0.78) & (-2.16, -1.28) & (-3.68, -2.11) & (33.6, 46.1) \\ 
  C2 & -0.03 & -0.71 & 0.19 & 0.68 & -1.77 & -0.95 & -2.72 & 65.2 \\ 
   & (-0.05, -0.01) & (-1.03, -0.41) & (0.17, 0.23) & (0.64, 0.72) & (-2.69, -1.02) & (-1.32, -0.58) & (-3.87, -1.58) & (60.1, 72.2) \\ 
  C3 & -0.10 & -2.20 & 0.35 &  &  & -2.20 & -2.20 &  \\ 
   & (-0.13, -0.07) & (-2.55, -1.82) & (0.32, 0.40) &  &  & (-2.55, -1.82) & (-2.55, -1.82) &  \\ 
  C4 & -0.02 & -0.45 & 0.18 & 0.81 & -1.54 & -0.60 & -2.13 & 71.6 \\ 
   & (-0.04,  0.00) & (-0.71, -0.16) & (0.16, 0.22) & (0.79, 0.83) & (-2.50, -0.61) & (-0.95, -0.26) & (-3.42, -0.95) & (65.2, 76.1) \\ 
  C5 & -0.03 & -0.70 & 0.19 & 0.69 & -1.80 & -0.94 & -2.73 & 65.6 \\ 
   & (-0.05, -0.01) & (-1.03, -0.37) & (0.16, 0.23) & (0.65, 0.73) & (-2.89, -0.94) & (-1.33, -0.55) & (-4.12, -1.52) & (60.6, 72.1) \\ 
\bottomrule
\end{tabular*}
\begin{tablenotes}[para,flushleft]
\tiny{\textit{Notes}: For model specifications see Table \ref{tab:models}. For those featuring heterogeneous coefficients over the cross-section $i$ or over time $t$, we take the arithmetic mean over all industries and over time per iteration of the algorithm and report the resulting posterior percentiles (the posterior median, and the bounds in parentheses marking the $99$ percent posterior credible set).}
\end{tablenotes}
\end{threeparttable}
\end{tiny}
\end{center}
\label{tab:results_split2002}
\end{table*}

\clearpage
\subsection{Split-sample analysis, 2001--2008 (last 60 FOMC meetings)}
\begin{table*}[htbp]
\caption{Estimated impacts of monetary policy on stock returns across industries, 1992 IO-tables.}\vspace*{-1.5em}
\begin{center}
\begin{tiny}
\begin{threeparttable}
\begin{tabular*}{\textwidth}{@{\extracolsep{\fill}} lcccccccc}
\toprule
               & \multicolumn{4}{c}{\textbf{Parameters}} & \multicolumn{4}{c}{\textbf{Impacts}}\\
               \cmidrule(lr){2-5}\cmidrule(lr){6-9}
   & $\alpha$ & $\beta$ & $\sigma^2$ & $\rho$ & Indirect & Direct & Total & Netw. (\%) \\ 
\midrule
A1 & -0.15 & -4.04 & 0.27 &  &  & -4.04 & -4.04 &  \\ 
   & (-0.34,  0.03) & (-7.28, -0.91) & (0.17, 0.48) &  &  & (-7.28, -0.91) & (-7.28, -0.91) &  \\ 
  A2 & -0.17 & -3.63 & 0.22 &  &  & -3.63 & -3.63 &  \\ 
   & (-0.32, 0.02) & (-6.54, -0.89) & (0.14, 0.41) &  &  & (-6.54, -0.89) & (-6.54, -0.89) &  \\ 
   \midrule
  B1 & -0.14 & -4.27 & 0.08 &  &  & -4.27 & -4.27 &  \\ 
   & (-0.16, -0.12) & (-4.63, -3.91) & (0.07, 0.08) &  &  & (-4.63, -3.91) & (-4.63, -3.91) &  \\ 
  B2 & -0.14 & -4.01 & 0.32 &  &  & -4.01 & -4.01 &  \\ 
   & (-0.16, -0.12) & (-4.42, -3.66) & (0.30, 0.34) &  &  & (-4.42, -3.66) & (-4.42, -3.66) &  \\ 
  B3 & -0.02 & -0.59 & 0.06 & 0.87 & -3.49 & -0.90 & -4.40 & 79.4 \\ 
   & (-0.03, -0.01) & (-0.77, -0.43) & (0.06, 0.07) & (0.85, 0.88) & (-4.39, -2.53) & (-1.16, -0.66) & (-5.47, -3.20) & (77.3, 81.3) \\ 
  B4 & -0.01 & -0.40 & 0.06 & 0.91 & -3.62 & -0.65 & -4.28 & 84.7 \\ 
   & (-0.02,  0.00) & (-0.55, -0.24) & (0.05, 0.06) & (0.90, 0.92) & (-5.10, -2.12) & (-0.87, -0.42) & (-6.02, -2.55) & (82.0, 87.1) \\
   \midrule 
  C1 & -0.02 & -0.58 & 0.06 & 0.74 & -2.65 & -0.83 & -3.48 & 76.1 \\ 
   & (-0.03, -0.01) & (-0.80, -0.36) & (0.06, 0.07) & (0.71, 0.77) & (-5.22, -1.65) & (-1.15, -0.52) & (-6.21, -2.17) & (73.2, 83.3) \\ 
  C2 & -0.02 & -0.41 & 0.06 & 0.81 & -2.79 & -0.61 & -3.41 & 82.0 \\ 
   & (-0.03, -0.01) & (-0.61, -0.22) & (0.06, 0.07) & (0.78, 0.83) & (-5.26, -1.50) & (-0.91, -0.35) & (-6.09, -1.87) & (78.2, 87.9) \\ 
  C3 & -0.14 & -4.00 & 0.32 &  &  & -4.00 & -4.00 &  \\ 
   & (-0.16, -0.12) & (-4.34, -3.65) & (0.30, 0.34) &  &  & (-4.34, -3.65) & (-4.34, -3.65) &  \\ 
  C4 & -0.01 & -0.38 & 0.05 & 0.91 & -3.49 & -0.67 & -4.15 & 83.9 \\ 
   & (-0.02,  0.00) & (-0.54, -0.21) & (0.05, 0.06) & (0.89, 0.92) & (-4.96, -2.18) & (-0.91, -0.39) & (-5.84, -2.57) & (81.3, 86.2) \\ 
  C5 & -0.01 & -0.58 & 0.06 & 0.82 & -4.57 & -0.95 & -5.52 & 82.7 \\ 
   & (-0.03, -0.01) & (-0.79, -0.37) & (0.05, 0.06) & (0.80, 0.84) & (-9.46, -2.61) & (-1.29, -0.63) & (-10.62, -3.25) & (78.6, 89.0) \\  
\bottomrule
\end{tabular*}
\begin{tablenotes}[para,flushleft]
\tiny{\textit{Notes}: For model specifications see Table \ref{tab:models}. For those featuring heterogeneous coefficients over the cross-section $i$ or over time $t$, we take the arithmetic mean over all industries and over time per iteration of the algorithm and report the resulting posterior percentiles (the posterior median, and the bounds in parentheses marking the $99$ percent posterior credible set).}
\end{tablenotes}
\end{threeparttable}
\end{tiny}
\end{center}
\label{tab:results_split21992}
\end{table*}

\begin{table*}[htbp]
\caption{Estimated impacts of monetary policy on stock returns across industries, 1997 IO-tables.}\vspace*{-1.5em}
\begin{center}
\begin{tiny}
\begin{threeparttable}
\begin{tabular*}{\textwidth}{@{\extracolsep{\fill}} lcccccccc}
\toprule
               & \multicolumn{4}{c}{\textbf{Parameters}} & \multicolumn{4}{c}{\textbf{Impacts}}\\
               \cmidrule(lr){2-5}\cmidrule(lr){6-9}
   & $\alpha$ & $\beta$ & $\sigma^2$ & $\rho$ & Indirect & Direct & Total & Netw. (\%) \\ 
\midrule
  B1 & -0.14 & -4.08 & 0.08 &  &  & -4.08 & -4.08 &  \\ 
   & (-0.16, -0.12) & (-4.41, -3.75) & (0.07, 0.08) &  &  & (-4.41, -3.75) & (-4.41, -3.75) &  \\ 
  B2 & -0.13 & -3.84 & 0.32 &  &  & -3.84 & -3.84 &  \\ 
   & (-0.15, -0.11) & (-4.15, -3.54) & (0.30, 0.34) &  &  & (-4.15, -3.54) & (-4.15, -3.54) &  \\ 
  B3 & -0.02 & -0.71 & 0.08 & 0.83 & -3.19 & -0.98 & -4.17 & 76.6 \\ 
   & (-0.03, -0.01) & (-0.89, -0.54) & (0.08, 0.09) & (0.81, 0.85) & (-3.96, -2.47) & (-1.22, -0.75) & (-5.17, -3.22) & (74.5, 78.4) \\ 
  B4 & -0.01 & -0.29 & 0.08 & 0.93 & -3.49 & -0.51 & -4.01 & 87.3 \\ 
   & (-0.02,  0.00) & (-0.45, -0.12) & (0.07, 0.08) & (0.91, 0.95) & (-5.29, -1.79) & (-0.72, -0.27) & (-5.96, -2.07) & (84.6, 89.9) \\ 
   \midrule
  C1 & -0.02 & -0.52 & 0.08 & 0.76 & -1.93 & -0.70 & -2.63 & 73.6 \\ 
   & (-0.03, -0.01) & (-0.73, -0.30) & (0.08, 0.09) & (0.74, 0.78) & (-2.68, -1.20) & (-0.97, -0.40) & (-3.64, -1.62) & (71.1, 76.3) \\ 
  C2 & -0.01 & -0.26 & 0.08 & 0.89 & -2.61 & -0.44 & -3.06 & 85.5 \\ 
   & (-0.02,  0.00) & (-0.45, -0.07) & (0.07, 0.08) & (0.87, 0.90) & (-4.55, -0.88) & (-0.70, -0.18) & (-5.22, -1.01) & (80.1, 89.2) \\ 
  C3 & -0.13 & -3.85 & 0.32 &  &  & -3.85 & -3.85 &  \\ 
   & (-0.15, -0.12) & (-4.20, -3.53) & (0.30, 0.33) &  &  & (-4.20, -3.53) & (-4.20, -3.53) &  \\ 
  C4 & -0.01 & -0.30 & 0.08 & 0.93 & -3.47 & -0.51 & -3.99 & 87.1 \\ 
   & (-0.02,  0.00) & (-0.48, -0.11) & (0.07, 0.08) & (0.91, 0.94) & (-5.28, -1.71) & (-0.75, -0.27) & (-6.02, -2.04) & (84.4, 89.3) \\ 
  C5 & -0.01 & -0.28 & 0.08 & 0.89 & -2.78 & -0.47 & -3.25 & 85.6 \\ 
   & (-0.02,  0.00) & (-0.46, -0.10) & (0.07, 0.08) & (0.87, 0.91) & (-4.66, -1.18) & (-0.71, -0.22) & (-5.33, -1.40) & (82.5, 88.8) \\ 
\bottomrule
\end{tabular*}
\begin{tablenotes}[para,flushleft]
\tiny{\textit{Notes}: For model specifications see Table \ref{tab:models}. For those featuring heterogeneous coefficients over the cross-section $i$ or over time $t$, we take the arithmetic mean over all industries and over time per iteration of the algorithm and report the resulting posterior percentiles (the posterior median, and the bounds in parentheses marking the $99$ percent posterior credible set).}
\end{tablenotes}
\end{threeparttable}
\end{tiny}
\end{center}
\label{tab:results_split21997}
\end{table*}

\begin{table*}[htbp]
\caption{Estimated impacts of monetary policy on stock returns across industries, 2002 IO-tables.}\vspace*{-1.5em}
\begin{center}
\begin{tiny}
\begin{threeparttable}
\begin{tabular*}{\textwidth}{@{\extracolsep{\fill}} lcccccccc}
\toprule
               & \multicolumn{4}{c}{\textbf{Parameters}} & \multicolumn{4}{c}{\textbf{Impacts}}\\
               \cmidrule(lr){2-5}\cmidrule(lr){6-9}
   & $\alpha$ & $\beta$ & $\sigma^2$ & $\rho$ & Indirect & Direct & Total & Netw. (\%) \\ 
\midrule
  B1 & -0.13 & -4.15 & 0.08 &  &  & -4.15 & -4.15 &  \\ 
   & (-0.15, -0.11) & (-4.49, -3.82) & (0.07, 0.08) &  &  & (-4.49, -3.82) & (-4.49, -3.82) &  \\ 
  B2 & -0.13 & -3.89 & 0.32 &  &  & -3.89 & -3.89 &  \\ 
   & (-0.15, -0.11) & (-4.28, -3.55) & (0.30, 0.34) &  &  & (-4.28, -3.55) & (-4.28, -3.55) &  \\ 
  B3 & -0.02 & -0.62 & 0.08 & 0.84 & -3.13 & -0.85 & -3.98 & 78.5 \\ 
   & (-0.03, -0.01) & (-0.81, -0.43) & (0.08, 0.09) & (0.83, 0.86) & (-4.04, -2.18) & (-1.12, -0.60) & (-5.15, -2.78) & (76.4, 80.5) \\ 
  B4 & -0.01 & -0.32 & 0.08 & 0.91 & -3.51 & -0.58 & -4.10 & 85.9 \\ 
   & (-0.02,  0.00) & (-0.50, -0.14) & (0.07, 0.08) & (0.90, 0.93) & (-5.12, -1.93) & (-0.82, -0.33) & (-5.88, -2.25) & (82.3, 88.3) \\ 
   \midrule
  C1 & -0.03 & -0.60 & 0.08 & 0.71 & -2.23 & -0.78 & -3.01 & 74.3 \\ 
   & (-0.04, -0.02) & (-0.85, -0.35) & (0.08, 0.09) & (0.68, 0.74) & (-3.45, -1.36) & (-1.09, -0.46) & (-4.49, -1.79) & (71.0, 78.7) \\ 
  C2 & -0.02 & -0.37 & 0.08 & 0.82 & -3.10 & -0.59 & -3.71 & 83.7 \\ 
   & (-0.03, -0.01) & (-0.59, -0.16) & (0.07, 0.08) & (0.79, 0.84) & (-7.65, -1.42) & (-0.88, -0.31) & (-8.32, -1.74) & (79.5, 91.5) \\ 
  C3 & -0.13 & -3.90 & 0.32 &  &  & -3.90 & -3.90 &  \\ 
   & (-0.15, -0.11) & (-4.23, -3.56) & (0.30, 0.33) &  &  & (-4.23, -3.56) & (-4.23, -3.56) &  \\ 
  C4 & -0.01 & -0.33 & 0.07 & 0.90 & -3.42 & -0.63 & -4.06 & 84.3 \\ 
   & (-0.02,  0.00) & (-0.52, -0.15) & (0.07, 0.08) & (0.89, 0.91) & (-4.81, -1.84) & (-0.89, -0.39) & (-5.71, -2.26) & (81.6, 86.5) \\ 
  C5 & -0.02 & -0.48 & 0.07 & 0.83 & -3.79 & -0.77 & -4.58 & 82.9 \\ 
   & (-0.03, -0.01) & (-0.68, -0.28) & (0.07, 0.08) & (0.81, 0.85) & (-6.08, -2.31) & (-1.04, -0.50) & (-7.01, -2.85) & (79.5, 88.1) \\ 
\bottomrule
\end{tabular*}
\begin{tablenotes}[para,flushleft]
\tiny{\textit{Notes}: For model specifications see Table \ref{tab:models}. For those featuring heterogeneous coefficients over the cross-section $i$ or over time $t$, we take the arithmetic mean over all industries and over time per iteration of the algorithm and report the resulting posterior percentiles (the posterior median, and the bounds in parentheses marking the $99$ percent posterior credible set).}
\end{tablenotes}
\end{threeparttable}
\end{tiny}
\end{center}
\label{tab:results_split22002}
\end{table*}

\end{appendices}
\end{document}